\documentclass[preprint,10pt,3p]{elsarticle}
\usepackage{lineno,hyperref}
\usepackage{version}
\usepackage{float} 
\usepackage{textcomp}
\usepackage{amsmath} 

\usepackage{graphicx,pslatex,float,color,array,amssymb,amsmath,euscript}
\usepackage{subcaption}
\usepackage{multirow}

\usepackage{gensymb}
\usepackage{siunitx}

\usepackage[english]{babel}
\usepackage[utf8]{inputenc}
\modulolinenumbers[5]

\usepackage{booktabs}

\definecolor{Blue}{rgb}{0.3,0.3,0.9}
\definecolor{Std}{rgb}{0.,0.,0.}
\def\revn#1{\color{Std}{#1}\color{Std}}

\def\rev#1{\color{Std}{#1}\color{Std}}
\journal{JECS}

\begin{document}

\begin{frontmatter}

%% Title, authors and addresses

%% use the tnoteref command within \title for footnotes;
%% use the tnotetext command for theassociated footnote;
%% use the fnref command within \author or \address for footnotes;
%% use the fntext command for theassociated footnote;
%% use the corref command within \author for corresponding author footnotes;
%% use the cortext command for theassociated footnote;
%% use the ead command for the email address,
%% and the form \ead[url] for the home page:
%% \title{Title\tnoteref{label1}}
%% \tnotetext[label1]{}
%% \author{Name\corref{cor1}\fnref{label2}}
%% \ead{email address}
%% \ead[url]{home page}
%% \fntext[label2]{}
%% \cortext[cor1]{}
%% \affiliation{organization={},
%%             addressline={},
%%             city={},
%%             postcode={},
%%             state={},
%%             country={}}
%% \fntext[label3]{}

\title{Thermal Barrier Coatings in burner rig experiment analyzed through LAser Shock for DAmage Monitoring (LASDAM) method}
%technique OU Blister driven delamination for Thermal Barrier Coating in burner rig experiment  OU Thermal Barrier Coatings in burner rig experiments involving the LAser Shock for DAmage Monitoring (LASDAM) method}
%\title{Thermal gradient and cooling rate influences on thermal barrier coatings damage rate for edge delamination designed by LAser Shock for DAmage monitoring (LASDAM) technique}%Effect of thermal gradients on columnar structured YSZ thermal barrier coatings delamination from initial defects introduced by LASAT method}

\author[add1,add2]{Lara Mahfouz}
\author[add1,cor]{Vincent Maurel}
\author[add1]{Vincent Guipont}
\author[add1]{Basile Marchand}
\author[add1]{Rami El Hourany}
\author[add2]{Florent Coudon}
\author[add3]{Daniel E. Mack}
\author[add3]{Robert Va\ss{}en}

\cortext[cor]{Corresponding author. Tel.: +33-1-60 76 30 03;  Fax: +33-1-60 76 31 50\\
E-mail address: vincent.maurel@minesparis.psl.eu}
\address[add1]{MINES Paris - PSL , MAT - Centre des Mat\'eriaux, CNRS UMR 7633, BP 87 91003 Evry, France}
\address[add2]{Safran Tech, Etablissement Paris Saclay, rue des Jeunes Bois-Châteaufort, 78772 Magny-les-Hameaux, France}
\address[add3]{Forschungszentrum Jülich GmbH, Institute of Energy and Climate Research (IEK-1), Jülich, Germany}

\begin{abstract}
%% Text of abstract
\rev{This study investigates failure mechanisms in a typical thermal barrier coating (TBC) system comprising an EB-PVD columnar top coat, an aluminide bond coat, and a Ni-based single crystal superalloy substrate, simulating gas turbine operating conditions using a burner rig. TBC degradation, initiated by interfacial defects from the LASAT method, was studied during thermal gradient cycling under fast and slow cooling. In-situ optical and infrared imaging, along with ex-situ SEM cross-sectional analysis, monitored failure mechanisms. The Laser Shock for Damage Monitoring (LASDAM) technique provided insights into gradient and cooling rate impacts on columnar TBC damage. Results showed significant effects of cooling rate on delamination and localized failure at blister sites, with LASDAM revealing significant overheating at damage sites. Analysis included full-field temperature and damage assessment, emphasizing blister-driven delamination under severe thermal gradients. Discussion focused on elastic stored energy effects, noting that fast cooling induced transient conditions where reversed temperature gradients increased damage, limiting TBC lifespan}.

\end{abstract}

%Thermal cycling experiments applying a thermal gradient + LASAT + Numerical modeling


\end{frontmatter}

%% \linenumbers

%% main text

\section{Introduction}
\label{sec:intro}

A standard Thermal Barrier Coating (TBC) system comprises an underlying bond coat (BC), typically (Ni,Pt)Al, and an outer ceramic top coat (TC). In many industrial applications, yttria-stabilized zirconia (YSZ) is the preferred material for the top coat \cite{Evans:2008,maurel2022coated}. When in service at elevated temperatures, a thermally grown oxide (TGO), primarily alumina, naturally forms at the TC/BC interface. The interfacial stress arising from the thermal expansion coefficient (CTE) mismatch and the thickening of TGO during high-temperature exposure are widely acknowledged as failure mechanisms for EB-PVD thermal barrier coatings \cite{Evans:2008}. Previous studies have demonstrated that thermal cycling, coupled with growth strain from oxidation, induces rumpling and micro-cracking at the TC/BC interface \cite{Tolpygo:2000,Tolpygo:2001,Tolpygo:2004a}. It has been reported that the typical failure pattern for TBCs involves interfacial crack initiation at the TC/BC interface, progressing to macroscopic delamination. This is followed by the formation of a blister through the buckling of the ceramic top coat layer, ultimately leading to spallation failure of the TC \cite{Yanar:2006,Evans:2008,Courcier:2011}.

These findings underscore the necessity for robust experiments to investigate blister-driven delamination thoroughly \cite{scotson2024characterisation}. In line with this perspective, the LAser Shock for DAmage Monitoring (LASDAM) technique introduces new avenues for analyzing the driving forces behind delamination from an initial blister subjected to thermal cycling in a furnace \cite{Mahfouz:2023}. LASDAM involves inducing an artificial decohesion at the weak interface of the Thermal Barrier Coating (TBC) and monitoring its evolution during cycling. The former is achieved through precise control of the decohesion size and location using the LAser Shock Adhesion Testing (LASAT) technique \cite{boustie:2007}. The latter entails combining in-situ high-resolution CCD images, infrared thermography (IRT), and the measurement of 3D blister morphology. When applying the LASDAM method with thermal cycling in a furnace, it was observed that a significant initial increase in blister buckling precedes further delamination \cite{guipont:2019}. The driving forces for buckling were also identified as driving forces for this blister-driven delamination, a major intrinsic failure mode in TBCs \cite{Mahfouz:2023}.

The aforementioned failure mechanisms are primarily derived from assessments done with a uniform temperature field. However, real service conditions involve non-homogeneous temperature fields, where thermal gradients lead to different stress distributions in the TBC, resulting in varied damage behaviors compared to uniform temperature fields \cite{levi2012environmental, hutchinson2002delamination}. Burner rig cycling enhances the representativeness of thermal cycling testing. In contrast to traditional furnace cycling, burner rig testing enables the introduction of a thermal gradient throughout the thermal barrier system, along with a surface temperature gradient. This closely simulates service conditions, where combustion gases, cooling channels, and the low thermal conductivity of the ceramic induce a significant gradient of about \SI{100}{\celsius} through the coating thickness \cite{padture2002,EvansHE:2011}.

Most burner rig tests for YSZ TBCs have been performed on air plasma spray (APS) top coats and show as a notable result that the temperature gradient associated with the low conductivity of APS leads to partial spallation of the top coat \cite{li2017correlation}. However, the oxidation prior to the burner rig test shows a transition from cracking within the TC layer to interfacial cracking, depending on the burner rig cycling conditions and YSZ thickness, promoting interfacial decohesion \cite{vassen2000}. In addition, at elevated temperatures within a temperature gradient, some authors have reported sintering of the surface of the top layer, increasing both its conductivity and local Young's modulus \cite{kim2021method}. It is important to note that some studies dealing with APS TC assume that the presence of interfacial cracks does not modify the temperature gradient across the TC \cite{wu2019laser}, while others have demonstrated that a small temperature difference between crack faces results in a significant change in the mechanical state of the TBC \cite{qian1998effects,hutchinson2002delamination}. 

For EB-PVD YSZ top coats, damage is initiated by interfacial decohesion under both homogeneous temperature and burner rig conditions \cite{sniezewski2009thermal}. This same study indicates that thermal gradient cycling requires a greater number of cycles to induce failure compared to uniform temperature cycling, given a similar Thermal Growth Oxide (TGO) or interfacial temperature. The lower this temperature, the more pronounced the difference in experimental lifetime. However, these findings are influenced by substantial variations in cooling rates that could impact subsequent lifetime, coupled with edge effects driving Thermal Barrier Coating (TC) failure \cite{levi2012environmental}.

In addition to through-thickness temperature gradients, the surface temperature field on the turbine blade exhibits significant variability, creating surface thermal gradients that have been largely overlooked in the existing literature. Numerical studies, for instance, reveal temperature fluctuations exceeding 200 K over a distance of less than 1 mm, particularly near cooling holes \cite{jiang2019numerical}. Therefore, enhancing our comprehension of thermal barrier system behavior necessitates considering the effects of both through-thickness and in-plane thermal gradients.

This study aims to elucidate the local effects of through-thickness and in-plane temperature gradients on a typical Thermal Barrier Coating (TBC) system with a columnar EB-PVD top coat. For the first time, the LASDAM method is introduced to the burner rig setup and plays a central role in the analysis. The paper begins with a detailed description of the burner rig configuration and the LASDAM test conditions. The burner rig setup allows for the adjustment of the cooling rate. The analysis of local temperature, microstructure evolution and damage rate for two specified cooling conditions is then discussed. The results are compared to homogeneous temperature furnace cycling tests, and the subsequent discussion attempts to elucidate the extent to which local temperature and temperature gradient influence interfacial damage rates.

%The objective of this study is to gain knowledge on the local effect of through-thickness and in-plane temperature gradient of a typical TBC system for EB-PVD columnar top coat. The paper introduces for the first time LASDAM method to the burner rig setup. The contribution of LASDAM method is central to the proposed analysis.  First, the burner rig setup and LASDAM testing conditions are detailed. The burner rig facility allows modification of the cooling rate. Subsequently, local temperature, microstructure evolution and damage rate are analyzed for two prescribed cooling conditions. Results are compared to homogeneous temperature furnace cycling tests and the discussion aims to clarify the extent to which local temperature and temperature gradient control interface damage rate. 
%In this study, thermal gradient cycling behavior of the coatings was tested on a burner rig facility operating with a natural gas/oxygen mixture, at the IEK-1 institute in JÃ¼lich, Germany.

\section{Experimental methods}
\label{sec:methods}

\subsection{Specimen}
\begin{figure}[H]
	\centering
		\includegraphics[width=0.55\textwidth]{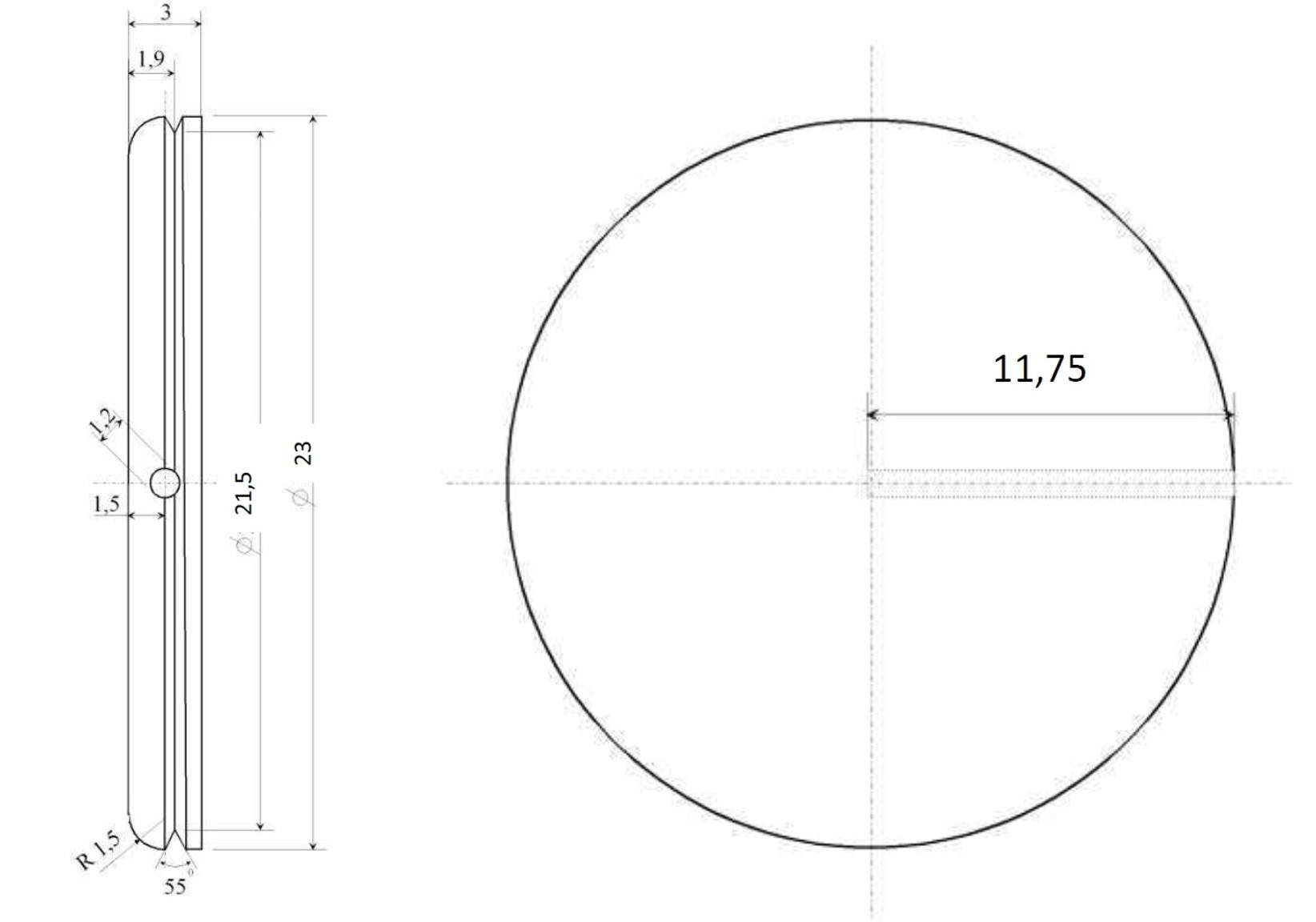}
	\caption {Samples geometry \rev{(dimensions are given in mm)}}
 \label{fig:plan}
\end {figure}

The studied TBC system is yttria (Y$_2$O$_3$) partially stabilized zirconia  (ZrO$_2$), YPSZ, ceramic EB-PVD top coat and (Ni,Pt)Al bond coat processed in the vapor phase with high temperature low activity  (APVS) treatment, deposited on a first generation Ni base single crystal superalloy AM1. Disk-shaped specimens, typical for burner rig tests, are employed in this study \cite{benoist2004microstructure, audigie2019high}. The superalloy substrate has a thickness of 3 mm and a diameter of 23 mm. In the as-processed state, the bond coat is approximately 50 µm thick, and the top coat is 145 ± 5 µm thick. To reduce stress levels and prevent failure at sharp edges, a 1.5 mm radius of curvature is machined at the outer edge of the samples. Additionally, a 1.2 mm hole is radially drilled in the substrate at mid-thickness up to the center of the sample, housing a thermocouple for measuring substrate temperature during burner rig cycling. A notch on the sample rim facilitates mounting in the sample holder of the burner rig test facility. Refer to Figure \ref{fig:plan} for a detailed view of the geometry. This study focuses on testing a single  batch of coating.

\subsection{Designing Interfacial Decohesion Using LAser Shock Adhesion Testing (LASAT)}
\label{sec:LASAT}
Interfacial cracks are processed within the coated samples by a laser shock using the LASAT method on a Saga Thales facility. This involves employing a Nd-YAG laser pulse with a wavelength $\lambda$=\SI{532}{nm} and a maximum energy of 2 J, lasting 5.2 ns. The LASAT process is preceded by 100 thermal cycles from 100 to \SI{1100}{\celsius}, each with a 50-minute dwell time at high temperature in a cyclic oxidation furnace at Safran \cite{Mahfouz:2023}. This pre-aging, consistent with \cite{Mahfouz:2023}, aims to initiate localized decohesion at the TC/TGO interface after the laser shock.

The laser shock diameter on the substrate surface is adjusted using a converging lens, and confinement is achieved by covering the substrate side with transparent adhesive tape. The laser power density ranges from 0.1 to 10 GW/cm², with laser diameters typically between 1 and \SI{4}{mm}; in this study, a 3 mm laser diameter was consistently used.

\begin{figure}[H]
	\centering
 \begin{subfigure}[b]{0.45\textwidth}
     \centering
     \includegraphics[width=\textwidth]{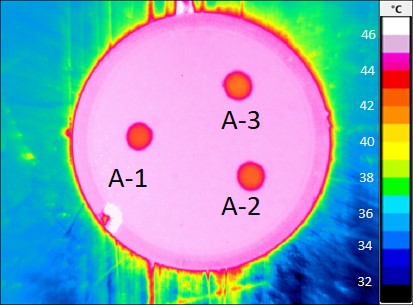}
     \caption{IRT after LASAT (sample A)}
 \label{fig:IR}
 \end{subfigure}
 \begin{subfigure}[b]{0.45\textwidth}
      \centering
	 \includegraphics[width=\textwidth]{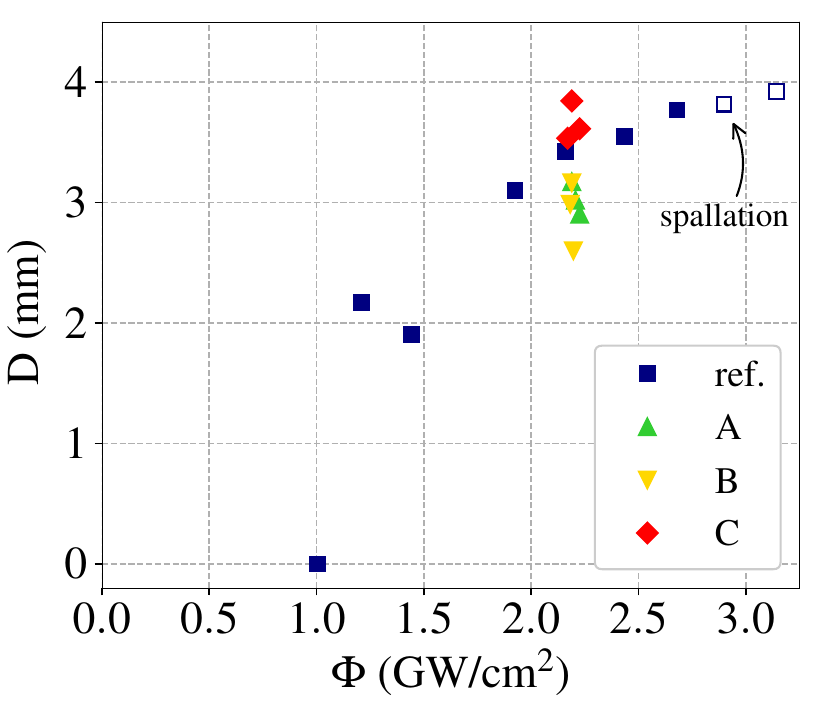}
      \caption{LASAT curve}
 \label{fig:LASAT}
 \end{subfigure}
 \begin{subfigure}[b]{0.55\textwidth}
      \centering
	 \includegraphics[width=\textwidth]{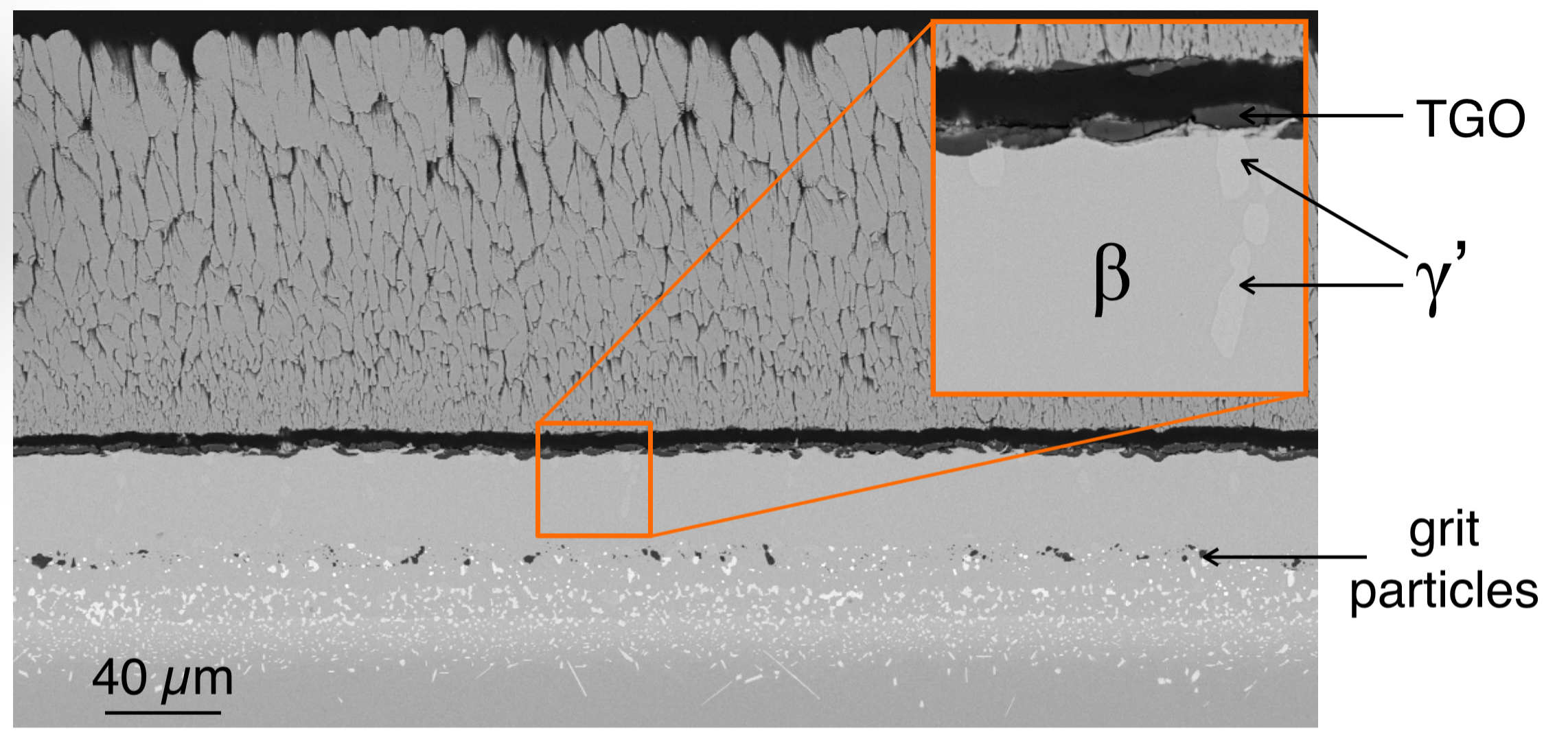}%w-eps-converted-to.pdf}
      \caption{SEM cross-section (Ref. sample)}
 \label{fig:SEM_LASAT}
 \end{subfigure}
	\caption{LASAT method: (a) Infrared thermography (IRT) employed for assessing the debonded diameter after laser shock   (b) Equivalent debonded diameters derived from IRT, with the arrow marking the initial occurrence of local spallation (full symbols denote perfect blisters, while empty symbols represent blisters with local or total spallation) \revn{(c) SEM cross section of a blister from the reference sample after LASAT test}}.
 \label{fig:IRLASAT}
\end{figure}

To measure the initial size of the debonded area post-LASAT processing, the sample is placed on a heating plate at 60 °C, and the surface temperature is measured using infrared thermography (IRT) (see Figure \ref{fig:IR}). The IRT camera, an INFRATEC ImageIR® 8300 hp with zoom capability, provides an in-plane spatial resolution of approximately \SI{42}{\micro\meter}.

A preliminary assessment of the adhesion level of the TBC was conducted on a reference sample (ref.) by creating a "LASAT curve," illustrating the evolution of the measured crack diameter in relation to the laser power density. The laser power density was systematically increased by gradually raising the laser energy for a constant laser spot diameter (3 mm in this study). The reported LASAT diameter represents the equivalent diameter of the debonded area measured by IRT, as depicted in Figure \ref{fig:LASAT}. The LASAT curve (ref. in \ref{fig:LASAT}) provides insights into the debonding laser power density threshold, serving as a reference to determine the laser energy level for inducing defects in the test samples.

Moreover, optical observations allow for the examination of whether delamination leads to buckling of the TC and if the resulting blister exhibits cracking or experiences local/total spallation. Notably, for higher laser energies, a characteristic LASAT transition from a "perfect blister" (without cracking or spallation) to a cracked/spalled state has been identified for a given TBC. This transition can be discerned by plotting a LASAT curve \cite{sapardanis:2016}. The pointed arrow in Figure \ref{fig:LASAT} emphasizes this transition, indicating the onset of spallation, with full symbols denoting perfect blisters and empty ones signifying local or total spallation. \rev{The SEM cross section of a blister from the reference sample used to plot the "LASAT curve" is shown in figure \ref{fig:SEM_LASAT}. It shows significant oxidation, oxide thickness locally reaching \SI{2}{\micro\meter}, and phase transformation from $\beta$ to $\gamma'$ phase, along with decohesion occurring at the interface between the oxide layer and the top coat layer, consistently with previous results \cite{guipont:2019}. The localization of the decohesion induced by LASAT confirms that the pre-aging weakens the oxide/top coat interface properties, consistent with the state-of-the-art for such TBC systems}.

We chose to introduce three defects on each burner rig test specimen, strategically avoiding critical areas that might interfere with the LASAT test, specifically, above the drilled hole and at the laser-engraved specimen reference. The application points for laser shocks were deliberately spaced apart to prevent potential interaction or premature coalescence. Laser shocks were applied at a target laser energy of 900 mJ, corresponding to a laser power density of approximately 2.2 {GW/cm$^2$}. The decohesion areas achieved on one of the samples (A) are shown in Figure \ref{fig:IR}. Equivalent diameters for all blisters introduced on test samples are plotted on the reference LASAT curve, as shown in Figure \ref{fig:LASAT}. While there is some variability in the resulting equivalent diameter among the samples, it can be considered negligible within the same sample. This variability is attributed to potential variations in TBC adhesion, even though the test samples are selected from a single batch. It is noteworthy that the LASAT technique has demonstrated high sensitivity to initial adhesion \cite{sapardanis:2017,maurel:2019}.

\begin{table}[!h]
    \centering
    \caption{Details of specimen tested, $\Phi$ is the applied shock energy, D$_0$ (mm) is the initial debonded diameter, $\Dot{T}$ is the average cooling rate, and N$_d$ and N$_s$  correspond respectively to the number of cycles for debonding and spallation when observed in situ. For the reference specimen used for LASAT curve, only ranges are indicated.}
    \begin{tabular}{ccc ccc c}
         sample & Spot & $\Phi$ (GW/cm$^2$) & D$_0$ (mm) & $\Dot{T}$ (\celsius.s$^{-1}$) & N$_d$ & N$_s$  \\
         \hline
         Ref. & 9 & 1-3.3 & 0-4 &-&-&-\\
         \cline{2-7}
        \multirow{3}{*}{A}&1&2.20&2.60&30&-&-\\
        &2&2.18&2.99&30&450&500\\
        &3&2.19&3.16&30&250&300\\
        \cline{2-7}
        \multirow{3}{*}{B}&1&2.23&2.90&30&-&-\\
        &2&2.21&3.02&30&490&500\\
        &3&2.19&3.17&30&290&300\\
        \cline{2-7}
        \multirow{3}{*}{C}&1&2.23&3.61&9&90&130\\
        &2&2.21&3.53&9&-&300\\
        &3&2.19&3.84&9&10&90\\
        
    \end{tabular}
    \label{tab:specimendetails}
\end{table}

\subsection{Burner Rig setup}
\begin{figure}[H]
	\centering
		\includegraphics[width=0.75\textwidth]{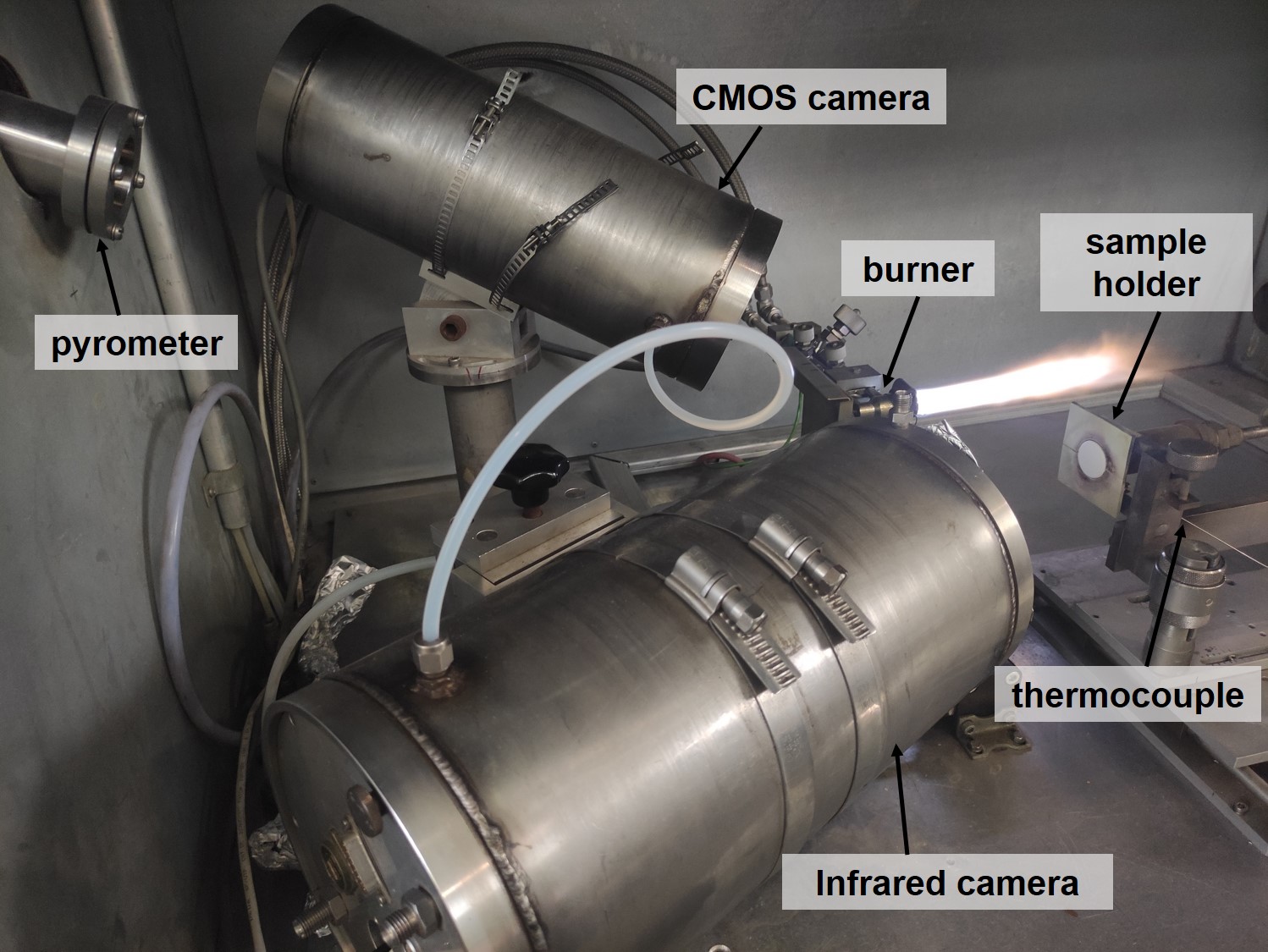}
	  \caption{Buner rig test set-up.}
 \label{fig:BR}
\end{figure}

The burner rig test facility utilized in this study is installed at IEK1 Julich. The set-up is detailed in Figure \ref{fig:BR}. The burner operates on a mixture of natural gas and O$_2$. 
During the heating phase, the burner flame is positioned over the ceramic surface, while the back side of the sample is cooled by compressed air to establish the desired temperature gradient across the thermal barrier coating. The temperature on the cold side of the sample is monitored using a thermocouple inserted into the substrate, while the temperature on the gas side of the sample is measured with a long-wave KT15.99 IIP (Heitronics\textregistered) pyrometer (9.6-11.5 $\mu$m). The recorded temperature value represents an average over a spot size of 10 mm in diameter focused on the center of the sample.

In this study, the temperature was set to \SI{1200}{\celsius} on the ceramic surface and \SI{1050}{\celsius} in the substrate. The coating's emissivity ($-eps-converted-to.pdfilon$) was assumed to be constant at 1. While this value is considered to underestimate the surface temperature by approximately 3\%, realistic emissivity values for YSZ ceramic in the pyrometer's wavelength range are around 0.98. However, emissions from the flame within the same spectral range partially counteract the impact of an overestimated emissivity \cite{Vassen2020}. \rev{Furthermore, it's important to note that the emissivity at the blisters might vary. This variability introduces uncertainty into the measurements, which is later addressed and considered in the analysis}.

Two cooling rates were examined. Fast cooling was achieved by moving the burner away immediately after the high-temperature dwell phase, while slow cooling was attained by maintaining the flame over the ceramic surface during cooling and reducing the gas flow rate. Details of the tested cycles will be provided in Section \ref{sec:temperature}.

A FLIR A655sc long-wave infrared (LWIR) infrared thermal camera with a resolution of 640 x 480 pixels and a spectral range of 7.5 - 14.0 $\mu$m provides in situ acquisition of the temperature field distribution on the surface of the ceramic, with a lateral resolution of about 63 $\mu$m. Simultaneously, an optical camera with a CMOS sensor (1400x1024 pixels) provides continuous high-resolution in situ observation of the coating surface. Both cameras are housed in cooled chambers during testing.

Intermittent pauses in cycling were implemented to use a CyberTECHNOLOGIES profilometer for measuring the 3D evolution of blisters. The profilometer has a lateral resolution of 3.5 $\mu$m and a vertical resolution of 35 nm. Cycling tests were stopped upon the observation of macroscopic restricted blister cracking, and samples were handled carefully to prevent additional damage to the top coat for subsequent analyses.

This study conducts a detailed examination of three distinct samples. Since three LASAT blisters are processed in each, a total of nine different blisters are tested using the proposed LASDAM method. The next section focuses on the thermal loading throughout burner rig cycling. Damage evolution in presence of the processed blister is later analyzed.

\section{Thermal loading experienced by blisters in burner rig cycling}
\subsection{Prescribed thermal cycles}
    \label{sec:temperature}

\begin{figure}[H]
	\centering
		
      \begin{subfigure}[b]{0.45\textwidth}
        \centering
        \includegraphics[width=\textwidth]{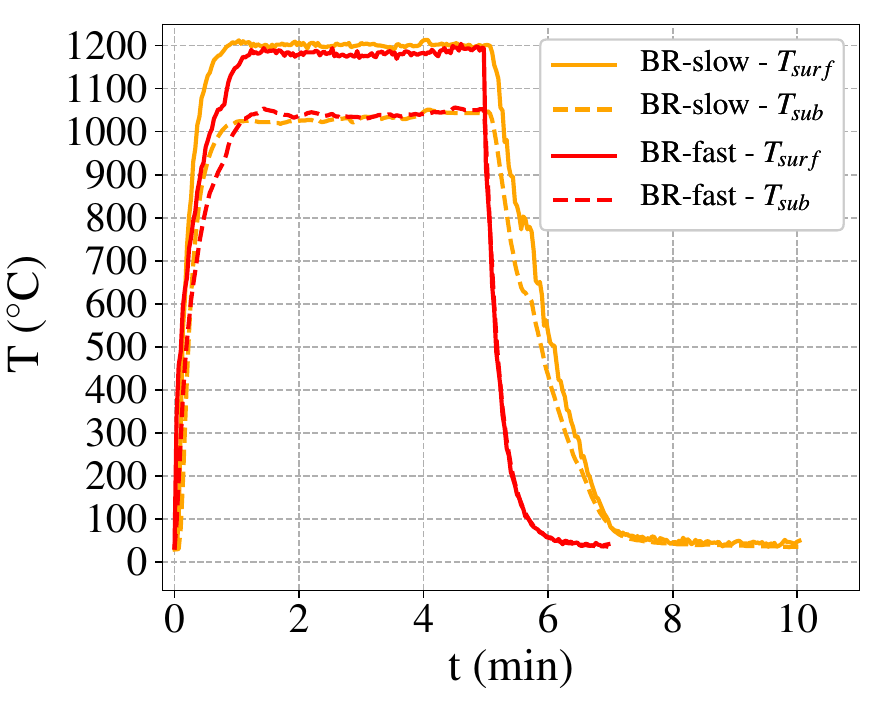}
 %       \caption{Pyrometer and TC\label{fig:cyclesT}}
        \caption{Temperature cycles at the ceramic surface ($T_{\text{surf}}$), represented by full lines, and in the substrate ($T_{\text{sub}}$), represented by dashed lines.}  \label{fig:cyclesT}
  \end{subfigure}
    \begin{subfigure}[b]{0.45\textwidth}
        \centering
        \includegraphics[width=\textwidth]{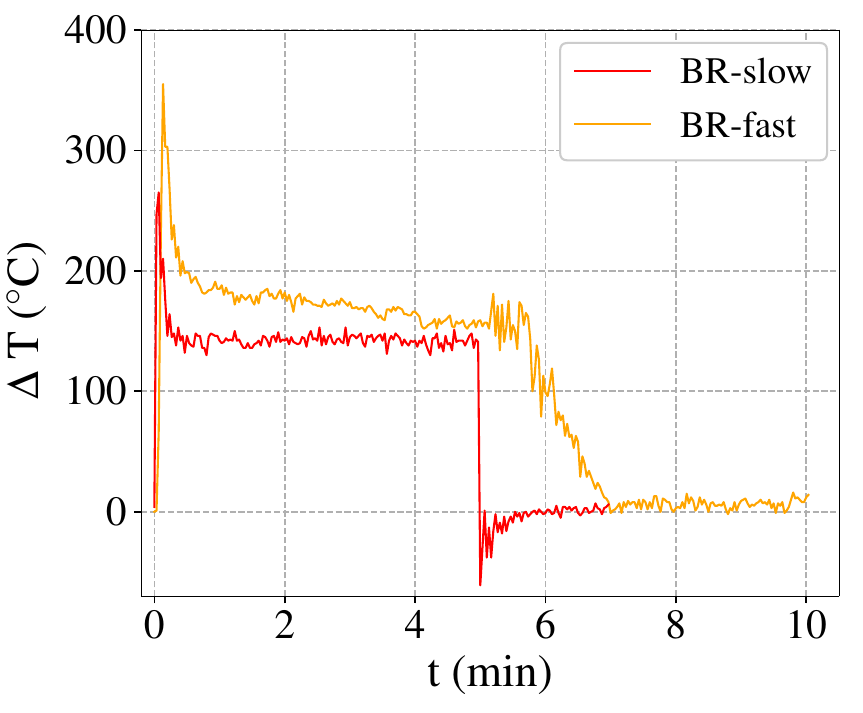}
        \caption{Through thickness temperature gradient\label{fig:cyclesDeltaT}} % $\Delta T_{exp}$} = T_{YSZ}-T_{AM1}$ 
    \end{subfigure}
		\caption{Temperature profiles recorded with a pyrometer on the ceramic surface and a thermocouple in the substrate during both fast and slow cooling rate cycles.} \label{fig:cycles}
\end{figure}

The burner rig cycles with slow and fast cooling rates are denoted as BR-slow and BR-fast, respectively. Figure \ref{fig:cycles} illustrates the temperature evolution over one cycle for both conditions. The surface temperature is obtained from the pyrometer measurement, while the substrate temperature is obtained from the thermocouple at mid-thickness in the substrate (see Figure \ref{fig:cyclesT}). Similar heating rates are observed in both conditions. The BR-fast cycle exhibits an average cooling rate of \SI{30}{\celsius/s} from the maximum substrate temperature to 200 $\celsius$, with a notably faster initial cooling rate. On the other hand, the BR-slow cycle has an average slope of 9 $\celsius$/s. The through-thickness temperature gradient is assessed as the temperature difference between the surface temperature measured with the pyrometer and the substrate temperature measured with the thermocouple at mid-thickness (see Figure \ref{fig:cyclesDeltaT}). Notably, the fast cooling rate cycle shows a reversal in the sign of temperature difference during the first few seconds of cooling, while the slow cooling rate cycle exhibits a slight increase in temperature difference during the initial seconds of cooling.

\subsection{Dwell at maximum temperature}
\label{sec:dwell_maxT}
\begin{figure}[H]
    \centering
    \begin{subfigure}[b]{0.42\textwidth}
        \centering
        \includegraphics[width=\textwidth]{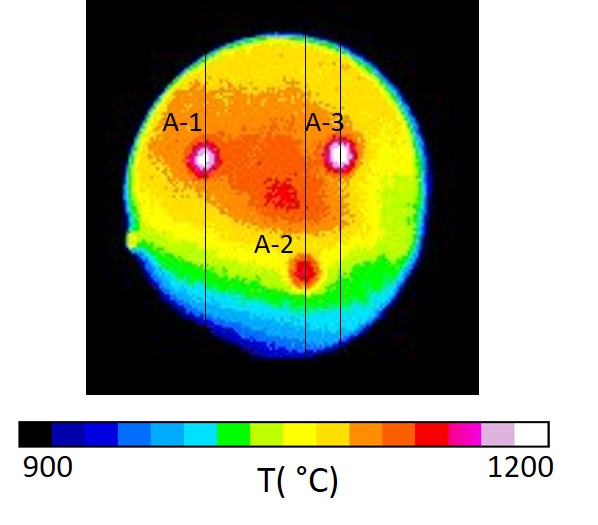}
        \caption{BR-slow (sample A) \label{fig:TfieldSlow}}
    \end{subfigure}
    \begin{subfigure}[b]{0.45\textwidth}
        \centering
        \includegraphics[width=\textwidth]{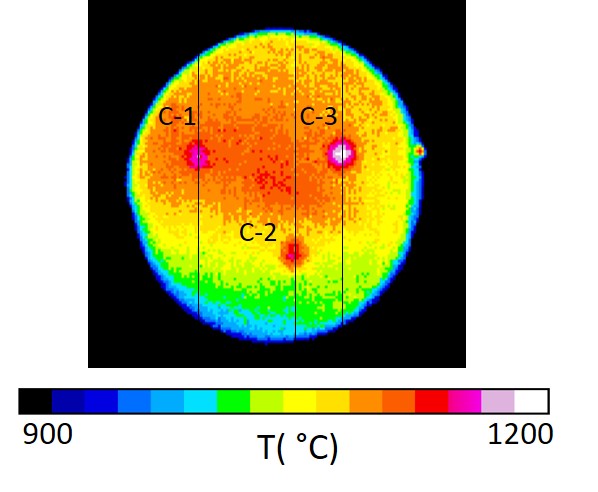}
        \caption{BR-fast (sample C) \label{fig:TfieldHigh}}
    \end{subfigure}
        \begin{subfigure}[b]{0.45\textwidth}
        \centering
        \includegraphics[width=\textwidth]{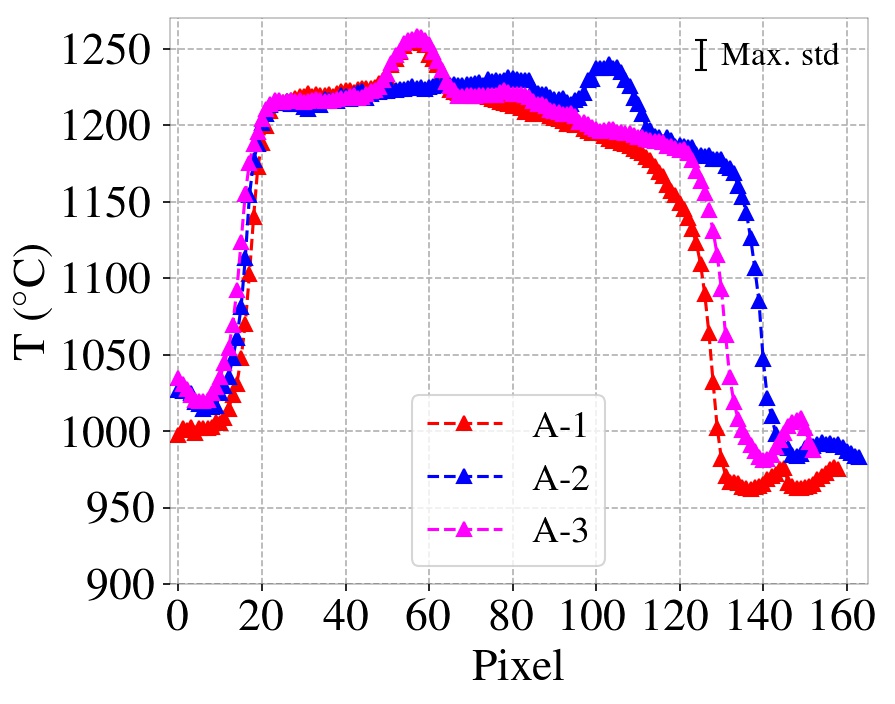}
        \caption{BR-slow (sample A)\label{fig:Tgrad29}}
    \end{subfigure}
    \begin{subfigure}[b]{0.45\textwidth}
        \centering
        \includegraphics[width=\textwidth]{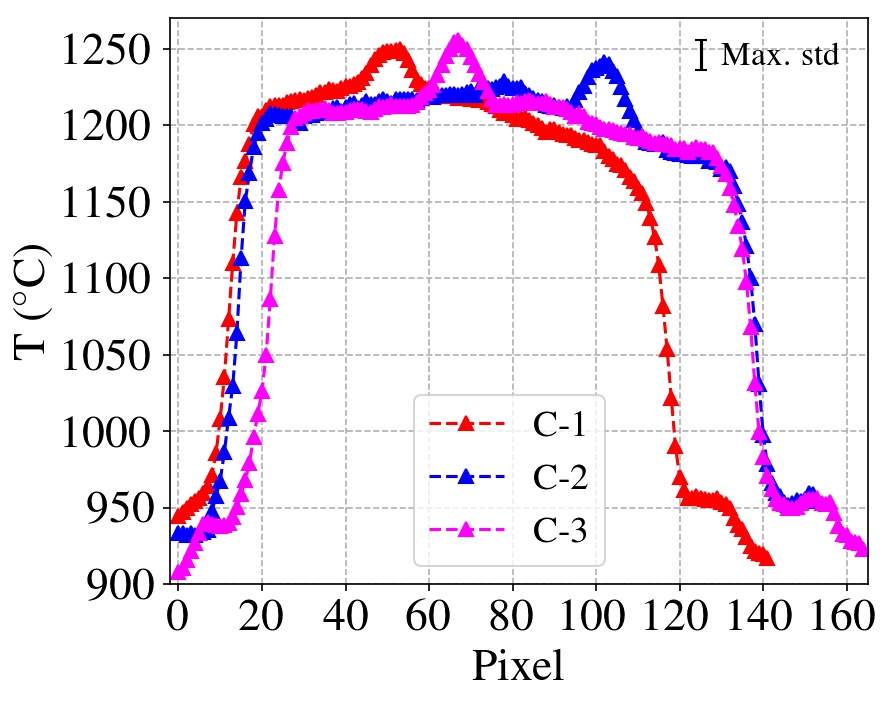}
        \caption{BR-fast (sample C) \label{fig:Tgrad19}}
    \end{subfigure}
    \caption{
   Temperature distribution captured by infrared thermography on the top coat surface during the dwell time at maximum temperature.}
    \label{fig:champsT}
\end{figure}

During the dwell time at maximum temperature, infrared thermography (IRT) captures the temperature distribution on the surface of the top coat, as shown in figures \ref{fig:champsT}(a) and (b). These observations are made after a few stabilization cycles for both samples A and C, tested with BR-slow and BR-fast, respectively. Throughout the sample, a central hot spot can be seen, due to the direct exposure of the surface of the top layer to the flame, with an off-center maximum temperature observed in the upper left corner of the infrared images. Additionally, local maximum temperatures are observed at blister locations.

Temperature profiles are generated along the black lines intersecting the blisters centers in Figures \ref{fig:TfieldSlow} and \ref{fig:TfieldHigh} for BR-slow and BR-fast conditions, respectively. The flame-associated temperature gradient is approximately 10 K, measured from 20 to 80 pixels on the temperature profile intersecting blister A-2  in Figure \ref{fig:Tgrad29}. Across blister A-2, the gradient is around 100 K, measured from 80 to 140 pixels (blue curve in Figure \ref{fig:Tgrad29}). In particular, a significant overheating of about 60 K is observed at the blister sites, extending from the centers to the edges of the blisters, as shown in Figures \ref{fig:Tgrad29} and \ref{fig:Tgrad19}.

Regardless of the maximum temperature recorded at the centers of the blisters, similar local temperatures are experienced at the edges of the blisters and their associated crack tips for both BR-fast and BR-slow conditions, approximately ranging from 1215 to $\SI{1220}{\celsius}$ at blister location \#3. At blister location \#2, corresponding to a lower temperature, both the local temperature and temperature gradient exhibit remarkable similarity for both conditions. These consistent observations affirm the robustness of the chosen experimental setup.

The observed overheating linked to each blister is likely a result of a layer of air beneath the blister, limiting the conduction of heat to the substrate. The maximum temperature observed in the center of a blister corresponds to the location of the maximum air gap thickness, while the gradual temperature decrease toward the edges of a blister corresponds to a continuous reduction in the air gap thickness, \rev{which raises questions about the influence of the blister's local height on overheating.}

\subsection{Temperature evolution of blisters during cycling}

\begin{figure}[H]
	\centering
     \begin{subfigure}[b]{0.45\textwidth}
        \centering
        \includegraphics[width=\textwidth]{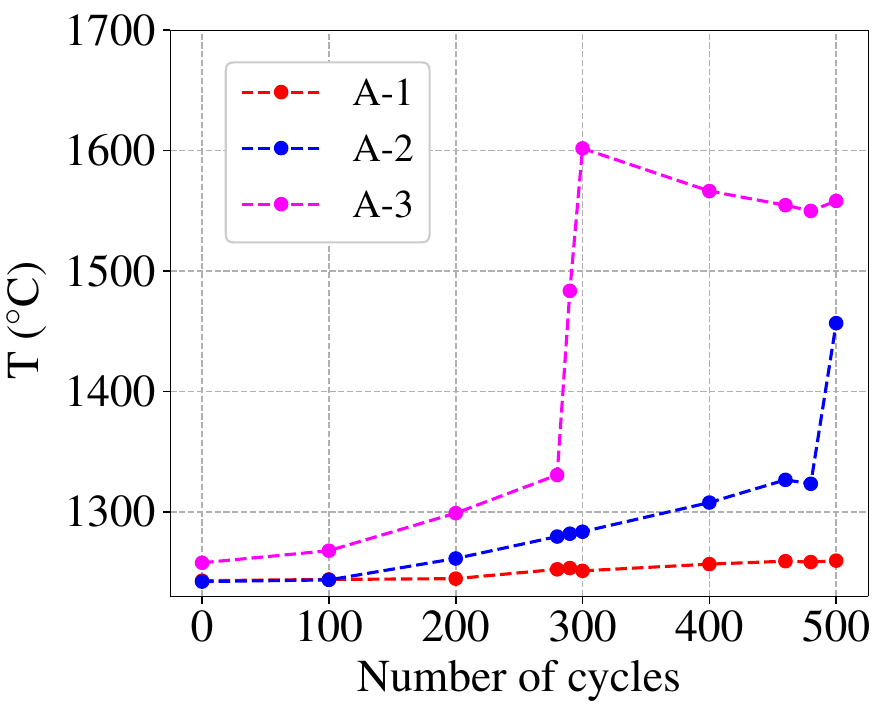}
        \caption{BR-slow (A)\label{fig:Tbl25N}}
    \end{subfigure}
    \begin{subfigure}[b]{0.45\textwidth}
        \centering
        \includegraphics[width=\textwidth]{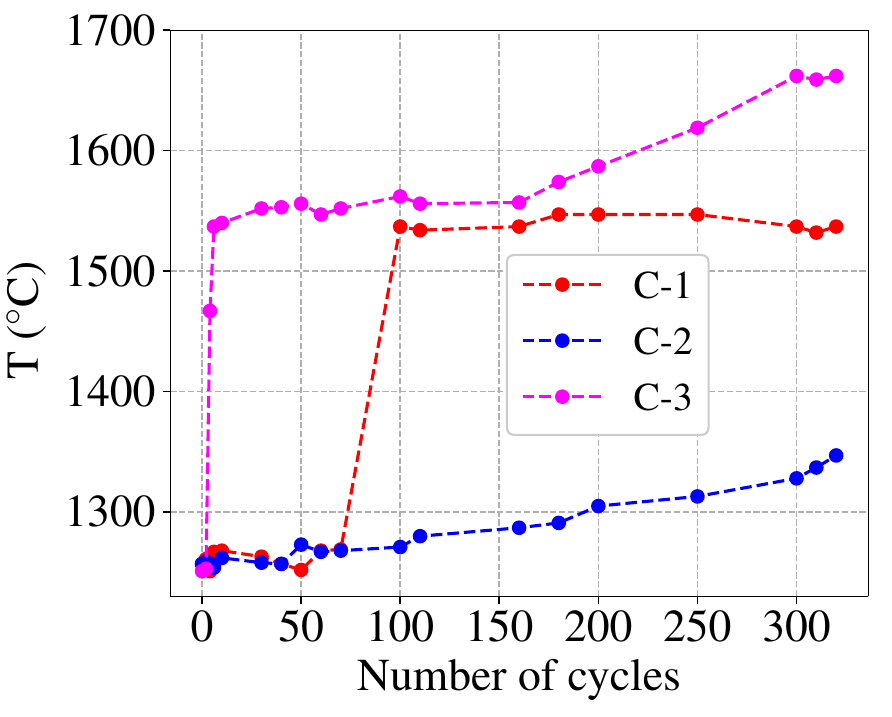}
        \caption{BR-fast (C) \label{fig:Tbl19N}}
    \end{subfigure}
	\caption{Evolution of local maximum temperatures at different blister locations over cycling.}
  \label{fig:T_cloques}
\end{figure}

Stable heating conditions are observed during cycling; nonetheless, there is a notable evolution in blisters temperature. In the case of BR-slow cycling, the maximum temperature at each of the three blisters exhibits a two-stage evolution. Initially, a continuous temperature increase is observed (e.g., up to N=280 for A-3 and N=470 for A-2, followed by a sudden temperature surge over a few cycles, as illustrated in Figure \ref{fig:T_cloques}(a). For A-1, only the first stage of evolution is evident, characterized by a slight temperature increase during cycling.

Similar two-step evolutions are observed for the BR-fast cycling condition. However, in the case of C-3, a drastic temperature increase occurs in fewer than 10 cycles.

Assuming that local overheating results from the presence of the air gap induced by blister buckling, the evolution of maximum temperatures indicates that the BR-fast cycling condition is more detrimental than the BR-slow cycling condition.  Blisters C-1 and C-3 are exposed to higher local temperatures than C-2 and experience a more rapid instability, a topic to be discussed in the \ref{sec:blisters} section.

\subsection{A focus on cooling condition}
\begin{figure}[H]
	\centering
     \begin{subfigure}[b]{0.45\textwidth}
        \centering
        \includegraphics[width=\textwidth]{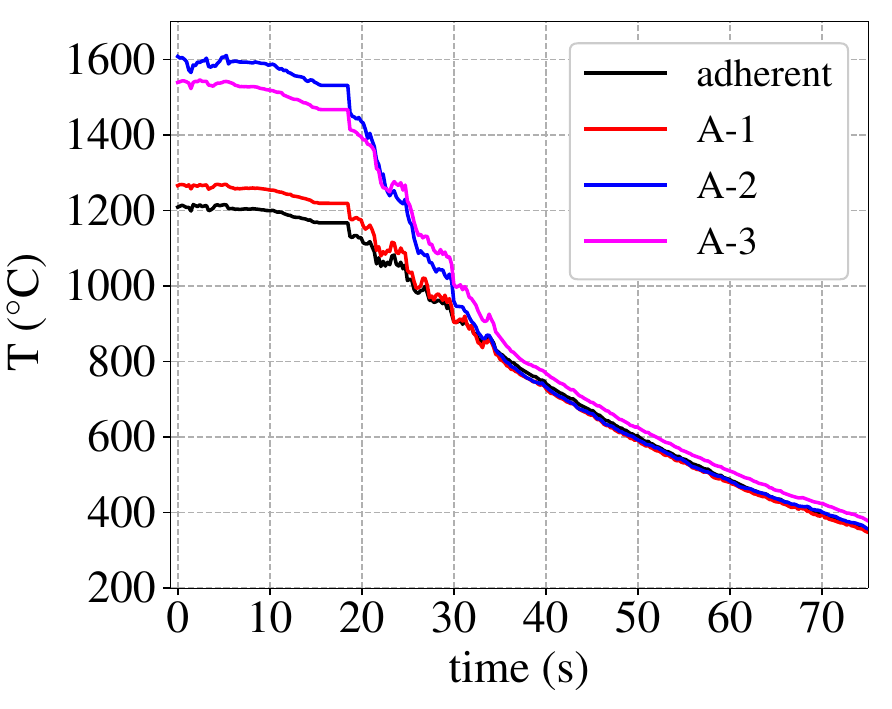}
        \caption{BR-slow (A) at N=500\label{fig:coolbl29N}}
    \end{subfigure}
    \begin{subfigure}[b]{0.45\textwidth}
        \centering
        \includegraphics[width=\textwidth]{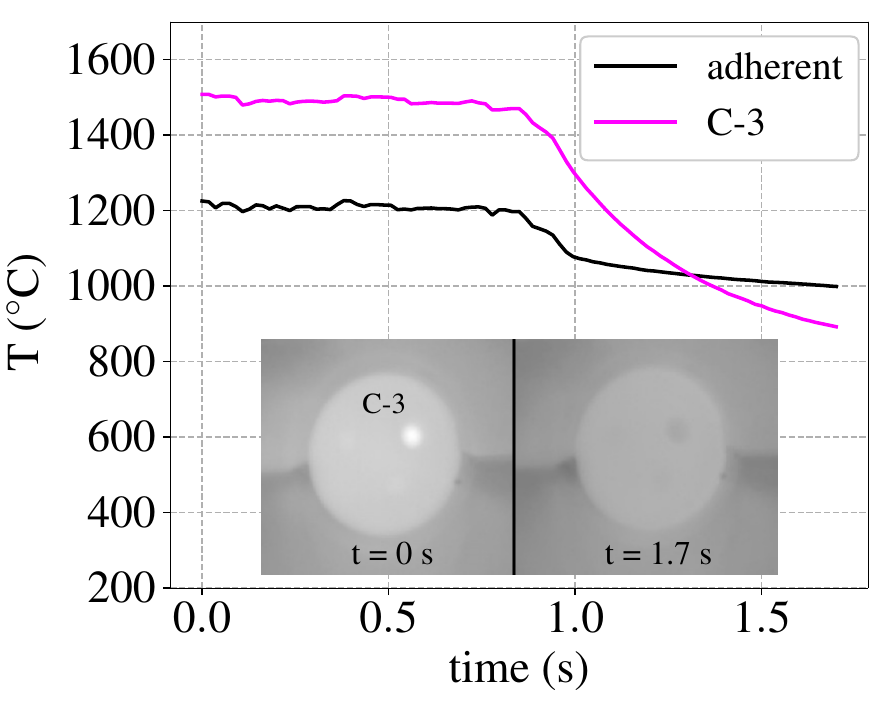}
        \caption{BR-fast (C) at N = 90\label{fig:coolbl19N}}
    \end{subfigure}
	\caption{Temperature profiles during cooling captured by infrared imaging at the blisters centers and a 10 mm circular reference area in the sample centers.}
  \label{fig:cooling_blisters}
\end{figure}

While the temperature range for IRT calibration applies to temperatures above \SI{400}{\celsius}, our focus is on cooling down to \SI{400}{\celsius}. Under the BR-slow cycling condition, achieved by decreasing the burner gas flow rate, Figure~\ref{fig:cooling_blisters}(a) illustrates that within less than 15 seconds, blister temperatures gradually align with the temperature measured at the adherent reference point. The temperature decreases from \SI{1530}{\celsius} to \SI{800}{\celsius}, considering the maximum temperature measured at A-2 after 490 cycles.

Conversely, for the BR-fast cycling condition, a similar temperature decrease is observed at cycle 10 in less than 1.5 seconds, as shown in Figure \ref{fig:coolbl19N}. Notably, the blister temperature drops below the temperature at the adherent reference point. This occurrence may be attributed to the sustained insulation effect of the air gap during cooling. The increased cooling rate, achieved by swiftly moving the burner away from the specimen can result in increased heat loss from the blister through direct radiation, surpassing the heat loss through conduction to the substrate in the adherent region. In the BR-slow condition, this effect is mitigated as the burner continues to heat the sample during the cooling phase, albeit at a reduced gas flow rate.

\section{Damage analysis}
\subsection{Evolution of blister morphology}
\label{sec:blisters}

\begin{figure}[H]
	\centering
		\includegraphics[width=0.95\textwidth]{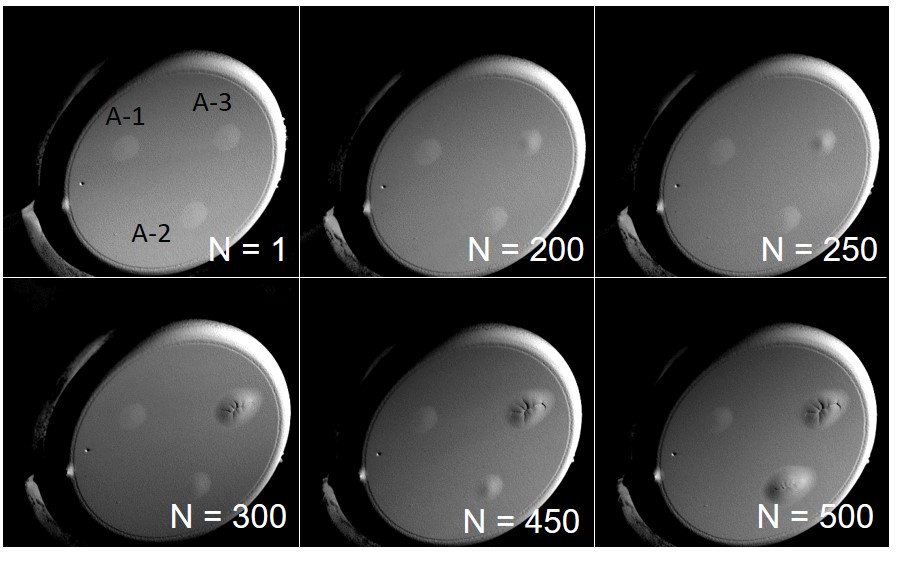}
	\caption{In situ CMOS analysis of samples subjected to BR-slow cooling rate (Sample A)}
  \label{fig:PC29in}
\end{figure}

In-situ imaging with the CMOS camera allows for a detailed analysis of the blister surface evolution throughout cycling. For the BR-slow condition (sample A), blister A-3, exposed to the maximum temperature, exhibits significant blistering after 200 cycles and develops cracks after 300 cycles (see Figure \ref{fig:PC29in}). In contrast, blister A-2, located in the lowest temperature region, displays substantial blistering after 450 cycles, and milder cracking becomes evident towards the end of the 500-cycle test. Blister A-1 remains more stable than the other two blisters. Cycling is stopped immediately upon the observation of cracking at blister A-2. Whether the damage observed at blister A-1 is related to initial lower adhesion at this location or to a higher in-plane temperature gradient requires further clarification by a more detailed analysis of the in-plane temperature gradient, possibly with a higher number of induced blisters within the sample area.

\begin{figure}[H]
	\centering
     \begin{subfigure}[b]{0.35\textwidth}
        \centering
        \includegraphics[width=\textwidth]{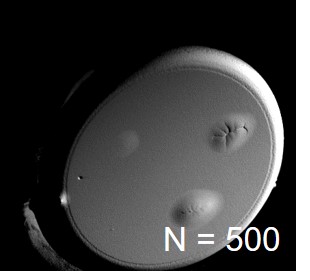}
        \caption {Sample A\label{fig:A}}
    \end{subfigure}
    \begin{subfigure}[b]{0.34\textwidth}
        \centering
        \includegraphics[width=\textwidth]{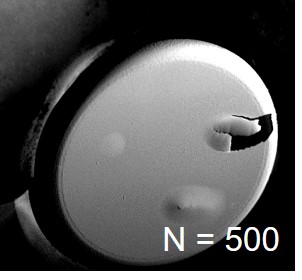}
        \caption{Sample B\label{fig:B}}
    \end{subfigure}
	\caption{In situ CMOS analysis of samples tested with the BR-slow cooling rate (samples A and B).}
  \label{fig:PC2925in}
\end{figure}

For the BR-slow condition, both tests conducted on samples A and B were stopped after 500 cycles, revealing a strikingly similar trend. The only exception was B-3, exhibiting spallation rather than severe cracking as observed for A-3 (Figure \ref{fig:PC2925in}). Clearly, the evolution of blisters is influenced by their specific location and associated maximum temperature. Notably, blister damage, whether in the form of cracking or spallation, is most significant at the location corresponding to blister \#3's maximum temperature. However, buckling and delamination are more pronounced at location \#2 compared to location \#1 on both samples A and B. This suggests that earlier cracking at site \#3 is associated with a maximum surface gradient (Figure \ref{fig:Tbl25N}).

\begin{figure}[H]
	\centering
		\includegraphics[width=0.95\textwidth]{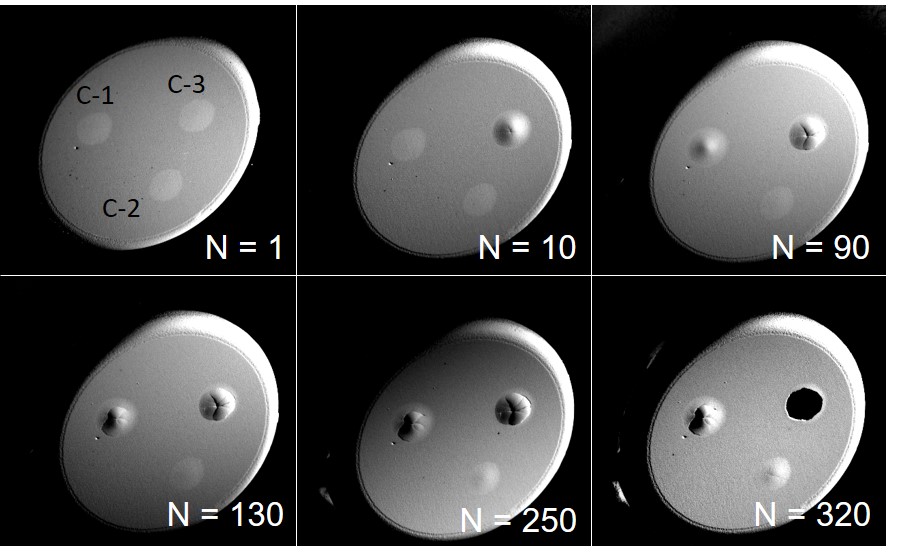}
	\caption{In situ CMOS captures at different number of cycles of the ceramic surface of a sample tested with BR-fast cooling rate (sample C).}
  \label{fig:PC19}
\end{figure}

In the BR-fast condition (sample C), blister C-3, exposed to the highest temperature, exhibits significant blistering after just 10 cycles, followed by cracking at 90 cycles and eventual spallation at 320 cycles (see Figure \ref{fig:PC19}). Meanwhile, the other two blisters display less blistering and no spallation at N=320. In particular, C-1, which is at a higher temperature, exhibits more pronounced blistering than C-2. In addition, C-1 exhibits surface roughness, which may be associated with localized cracking. The evolution of the blisters follows the observed temperature increase over the cycles, as shown in Figure \ref{fig:Tbl19N}. This figure highlights that the cooling rate determines the damage rate, with increased damage correlating with the maximum overheating of the blister.

%Thus, maximum heating drives the locus of maximum damage rate.

 \subsection{Microstructure evolution}

\begin{figure}[H]
	\centering
		\includegraphics[width=0.95\textwidth]{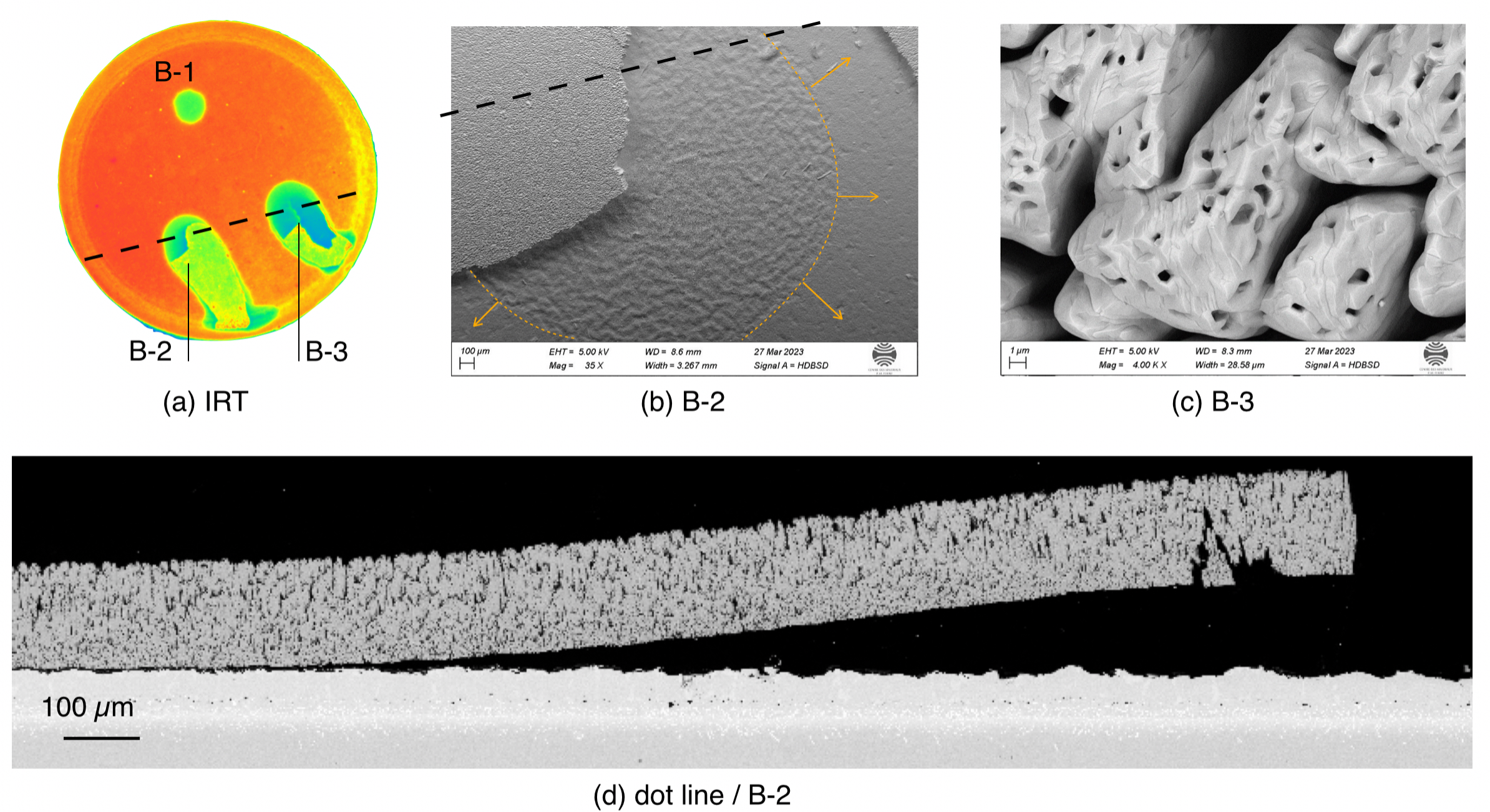}%PC25detailcomp3.jpeg}
	\caption{Ex situ analysis of a sample tested with BR-\rev{slow } cooling rate (sample B): \rev{(a) IRT observation of the surface of the specimen, (b) SEM surface observation of location B-2 orange dashed line corresponds to initial location of debonding processed by LASAT, orange arrows indicate further decohesion under burner rig cycling, (c) SEM surface observation of location B-3 and (d) SEM cross-section corresponding to the black dot lines in (a) and (b) at the location B-2. }}
  \label{fig:PC25ex}
\end{figure}

 \paragraph{BR-slow}
After testing, a higher-resolution ex situ Infrared Thermography (IRT) control of specimen B is conducted, as depicted in Figure \ref{fig:PC25ex}(a). The results affirm that the locations of damage align consistently with the maximum local temperatures at the blisters, and no other macroscopic defects are discernible. It is noteworthy that spallation on blister B-2, though not observed in situ, is evidently induced by a desktop spallation effect, as can be concluded from the comparison between Figures \ref{fig:PC25ex}(a) and \ref{fig:PC2925in}(b). A top-view scanning electron microscope (SEM) image in Figure \ref{fig:PC25ex}(c) reveals pronounced sintering of the top coat, highlighting the internal pores' aspect, and the absence of the typical feather-like structure.

The global surface view shown in Figure \ref{fig:PC25ex}(b), corresponding to blister B-2, shows the initial debonding site processed by LASAT, \rev{highlighted by the dashed line}. The maximum observed rumpling is confined within a circular region, clearly showing the onset of debonding. Beyond this region, \rev{one can observe the area where the interfacial crack has grown from this initial debonding site in the direction of the arrows. In this area of further debonding, } few local undulations are observed, consistent with established findings that rumpling manifests exclusively at points of local decohesion \cite{Mahfouz:2023}. Similarly, cross-sectional analysis of blister B-2, as shown in Figure \ref{fig:PC25ex}(d), confirms that the peak of rumpling occurs beneath the blister and is significantly reduced in regions where the topcoat adheres to the substrate. In addition, the adherent portion of the topcoat layer shows no signs of sintering, consistent with the lower surface temperature measured in this particular area, \rev{see Figure \ref{fig:Tgrad29}}.

\begin{figure}[H]
	\centering
		\includegraphics[width=0.95\textwidth]{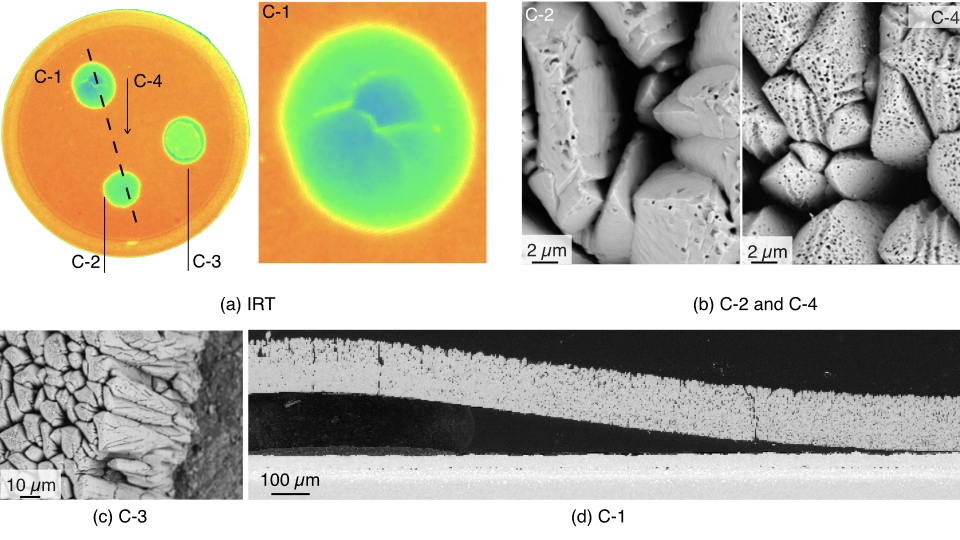}%}
	\caption{Ex situ analysis of a sample tested with BR-fast cooling rate (sample C): \rev{(a) IRT observation of the surface of the specimen, (b) SEM surface observation of location B-2 orange dashed line corresponds to initial location of debonding processed by LASAT, orange arrows indicate further decohesion under burner rig cycling, (c) SEM surface observation of location B-3 and (d) SEM cross-section corresponding to the black dot lines in (a) and (b) at the location B-2. }}
  \label{fig:PC19exc}
\end{figure}

 \paragraph{BR-fast}
 Examination conducted ex situ verifies that, under the BR-fast condition, blisters C-1 and C-2 remain unspalled. However, infrared thermography (IRT) reveals the presence of small cracks on blister C-1, as illustrated in Figure \ref{fig:PC19exc}(a). Meanwhile, spallation is complete for blister C-3, with no residual ceramic observed on the TGO surface. Additionally, sintering impacts the columnar structure of the top coat in the vicinity of blister C-3, as shown in Figure \ref{fig:PC19exc}(c). Notably, a feather-like structure persists at the coolest point, adjacent to adherent area C-4, where internal porosity is visible, as shown in Figure \ref{fig:PC19exc}(b). This porous structure is less prominent at other locations exposed to higher temperatures, illustrated in Figure \ref{fig:PC19exc}(b) for blister C-2.

The cross-sectional analysis along the line connecting blisters C-1 and C-2 in Figure \ref{fig:PC19exc}(a) reveals a noteworthy correlation between local high temperature during cycling and the sintering process, as depicted in Figure \ref{fig:PC19exc}(d). At the center of blister C-1, a highly dense microstructure is evident, indicative of localized extremely high temperatures. Importantly, the sintering is more pronounced in the lower part of the top coat (TC) throughout its thickness, consistent with local compressive stresses induced by the bending of the ceramic layer, \rev{but also corresponding to an initial denser part of the ceramic layer}.

Remarkably, at the location of the crack tip, and consistent with the lower local temperature, no discernible evidence of sintering is observed, as illustrated in Figure \ref{fig:PC19exc}(d). This observation underscores that, despite the elevated local temperature resulting from the blister-induced air gap, the crack propagates in an area characterized by lower and more uniform temperatures within the specimen's region of interest. In essence, the hot points induced by the air gap solely influence the temperature of the blisters, and blister driven delamination is analyzed at the targeted temperature.

More precise quantitative results are derived through cross-sectional image analysis. The average porosity is computed across a moving rectangular area, with a thickness matching that of the top layer, along the radial direction starting from the crack tip. The lateral dimensions of the rectangle are set to \SI{21}{\micro\meter}. To enhance the clarity of the obtained measurements, a moving average method is applied over 10 rectangles.

\begin{figure}[H]
    \centering
     \begin{subfigure}[b]{0.58\textwidth}
   \includegraphics[width=0.95\textwidth]{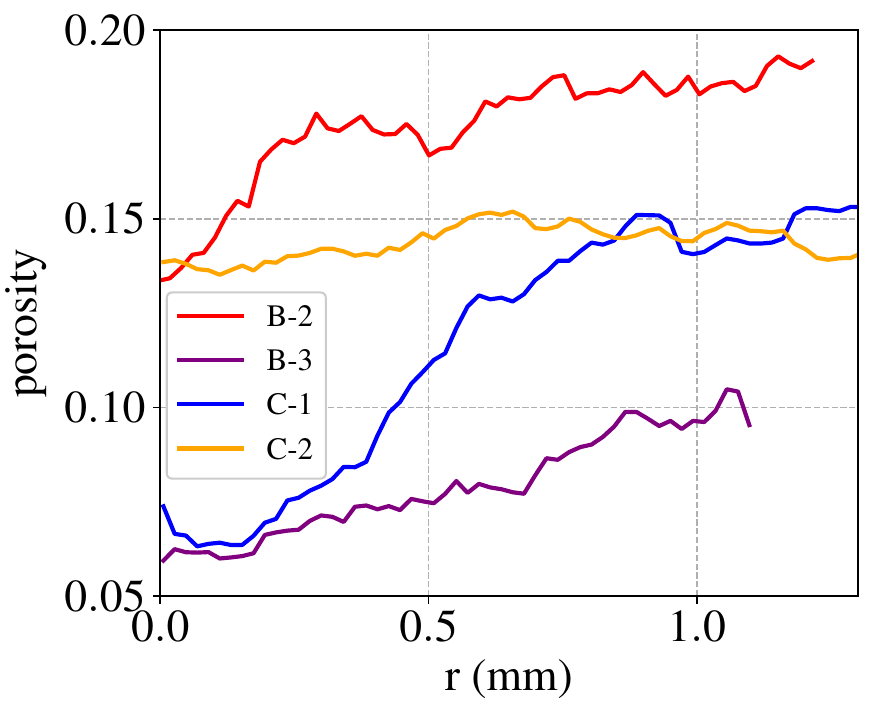} %IMG_6843.jpeg}%
	\caption{Porosity variation as a function of the distance from blister, denoted as r}
  \label{fig:porosity}
\end{subfigure}  
\begin{subfigure}[b]{0.42\textwidth}

        \centering
        \includegraphics[width=\textwidth]{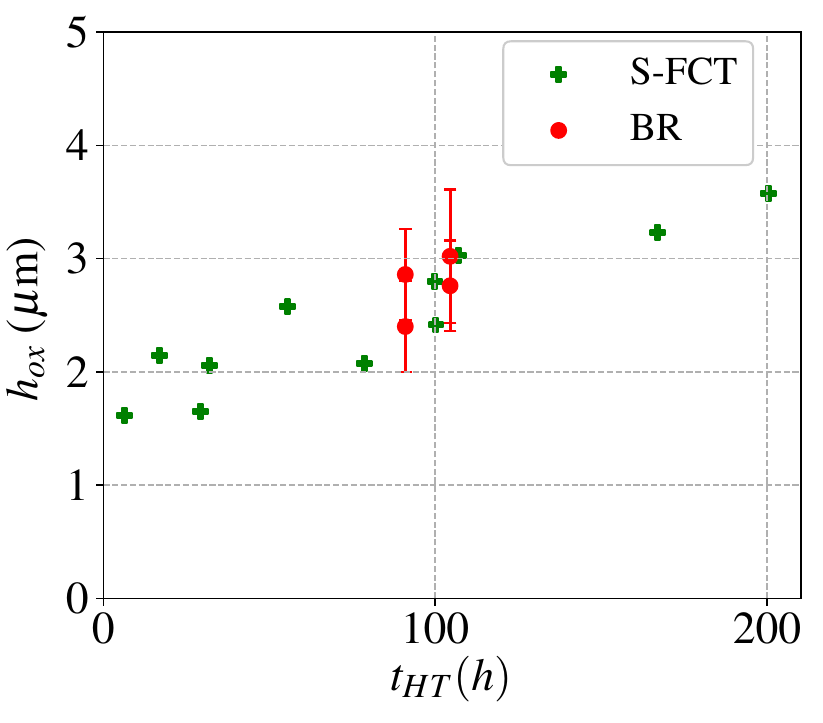}  
    \caption{Oxide thickness h$_{ox}$ \label{fig:eoxl}}
   \end{subfigure}
    \begin{subfigure}[b]{0.45\textwidth}
        \centering
        \includegraphics[width=\textwidth]{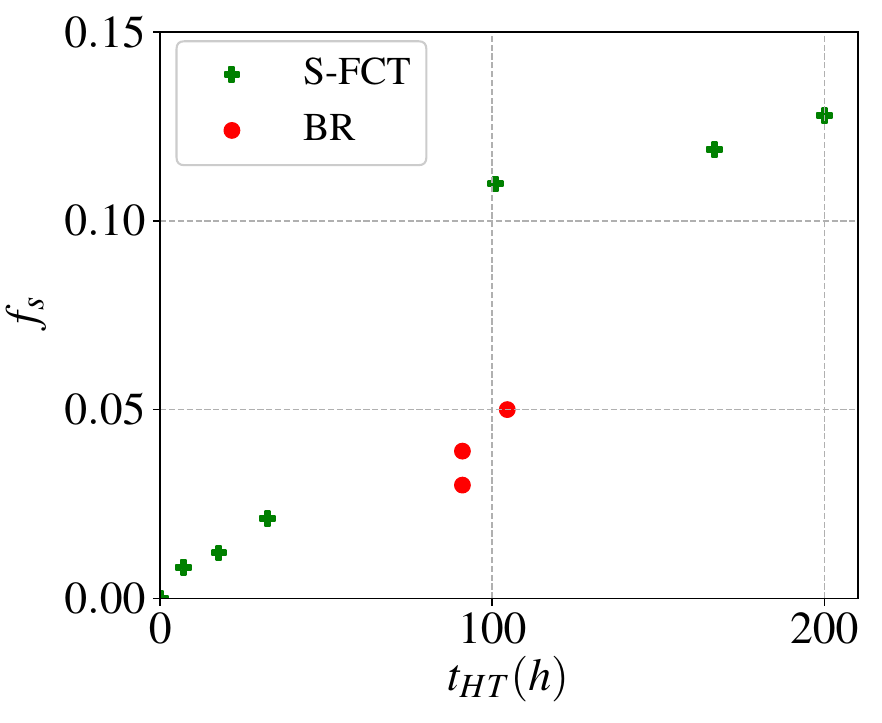}
       \caption{$\gamma$' phase fraction f$_s$\label{fig:Drel}}
\end{subfigure}

%refaire porosity avec traits plus fins et ticks plus petits
    \caption{Microstructure \rev{analysis considering (a) variation of porosity as a function of the distance from the center of the blister, (b) oxide thickness and (c) $\gamma$' phase fraction } evolutions for burner rig (BR/current study) and furnace cycling (S-FCT/data from \cite{soulignac2014prevision})}
  \label{fig:evolutionmic}
\end{figure}

It is evident that for blisters C-1 and B-3, the sintering is more pronounced in the center (resulting in lower porosity) and gradually decreases toward the crack tip (accompanied by an increase in porosity). In the case of sample C, where less time is spent at high temperatures compared to sample B, the porosity is expected to be inherently higher, as shown in Figure \ref{fig:porosity}. This is true for C-1 and C-2 compared to B-3. However, the measured porosity for B-2 is higher, possibly due to the slightly off-center cross section caused by the partial spallation of the blister, as shown in Figure \ref{fig:PC25ex}(d). It is noteworthy that the local temperature for B-2 may be lower than for C-2. Finally, it is important to note that given the inherent variability observed in the EB-PVD deposition process, the initial porosity may vary from one sample to another.

In addition to sintering of the top coat, SEM cross sections allow assessment of both local oxide thickness and phase transformation in the bond coat layer, from $\beta$-NiAl to $\gamma$'-Ni$_3$Al, Figures \ref{fig:evolutionmic}(b) and (c), respectively. \rev{Both were measured using SEM observations and segmentation through standard thresholding techniques using ImageJ/Fiji software \cite{ImageJ:2024}, for regions approximately \SI{1.4}{mm} wide in adherent top coat areas. The $\gamma'$ phase fraction corresponds to the surface fraction of the outer layer, defined as the area between the oxide layer and the embedded grit alumina particles, see Figure \ref{fig:SEM_LASAT}}.

Oxide thickness measurements were conducted for both BR-slow and BR-fast conditions on samples B (N=500, equivalent time at high temperature: 37.5 hours) and C (N=320, equivalent time at high temperature: 24 hours). These measurements were taken at two locations: the adherent part of the TC near the point used for pyrometer surface control and at the crack tip positions at B-2 for BR-slow and C-1 for BR-fast. Notably, B-2 exhibits the highest oxide thickness, possibly due to final overheating, while C-1 displays the lowest oxide thickness, likely attributed to local oxide spallation.

However, upon comparing oxide thickness below the blister and in areas where the TC is adherent, no significant difference in oxide thickness or phase transformation is observed (this corresponds to measurements at the same time but with different values). This observation suggests that, despite the elevated surface temperature caused by the blister, the variation in interfacial temperature can be considered small enough not to substantially impact the oxidation and phase transformation phenomena. A comparison with previous results obtained for S-FCT cycling will be discussed later.

In summary, the BR-fast condition induces spallation earlier than the BR-slow condition. The selected setup for the burner rig generates a non-uniform temperature field, and consequently the location of the blister with respect to the burner flame plays a crucial role in the damage induced. Furthermore, the presence of an air gap between the ceramic layer and the TGO surface promotes an increase in thermal resistance, consequently elevating the blister temperature and advancing sintering in the top coat.

\subsection{Analysis of damage rate}

\begin{figure}[H]
    \centering
    \begin{subfigure}[b]{0.45\textwidth}
        \centering
        \includegraphics[width=\textwidth]{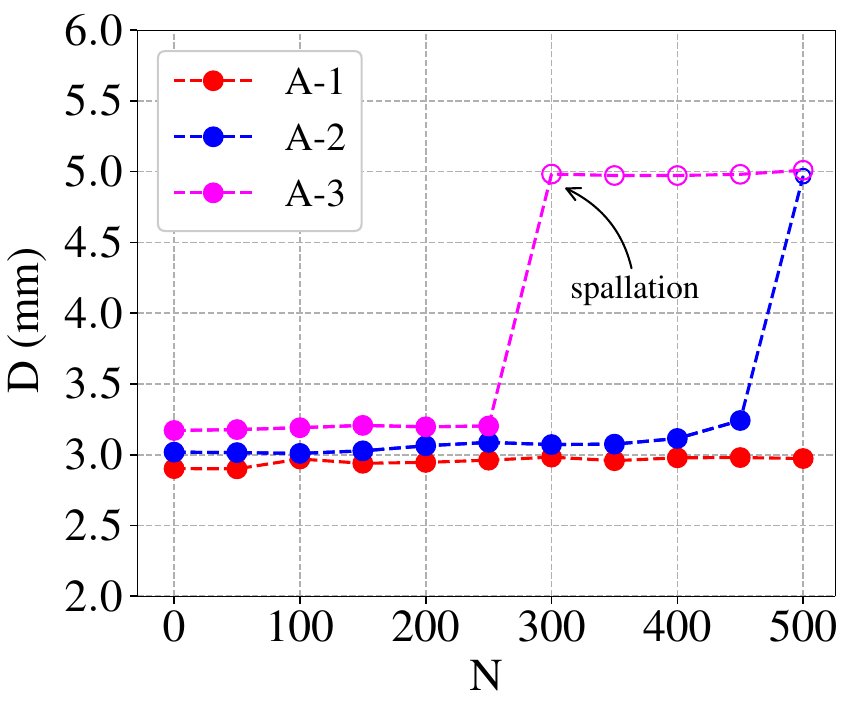}
        \caption{Debonded diameter \rev{sample A} \label{fig:DPC29}}
    \end{subfigure}
    \begin{subfigure}[b]{0.45\textwidth}
        \centering
        \includegraphics[width=\textwidth]{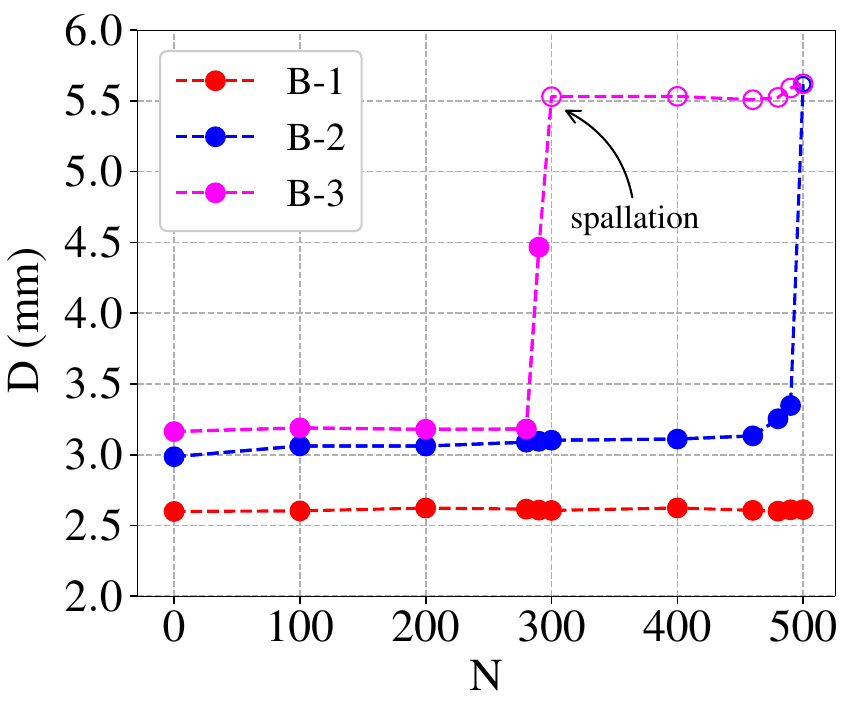}
        \caption{Debonded diameter  \rev{sample B} \label{fig:DPC25}}
    \end{subfigure}
    \begin{subfigure}[b]{0.45\textwidth}
        \centering
   		\includegraphics[width=\textwidth]{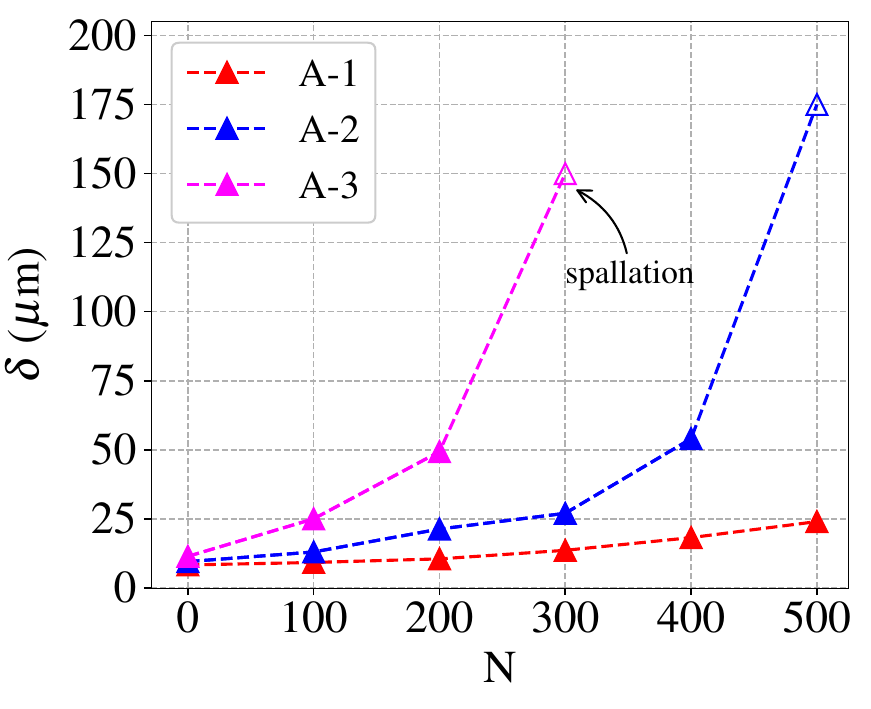}
	\caption{Blisters height \rev{sample A} }
  \label{fig:delta_PC29}
  \end{subfigure}
    \begin{subfigure}[b]{0.45\textwidth}
        \centering
   		\includegraphics[width=\textwidth]{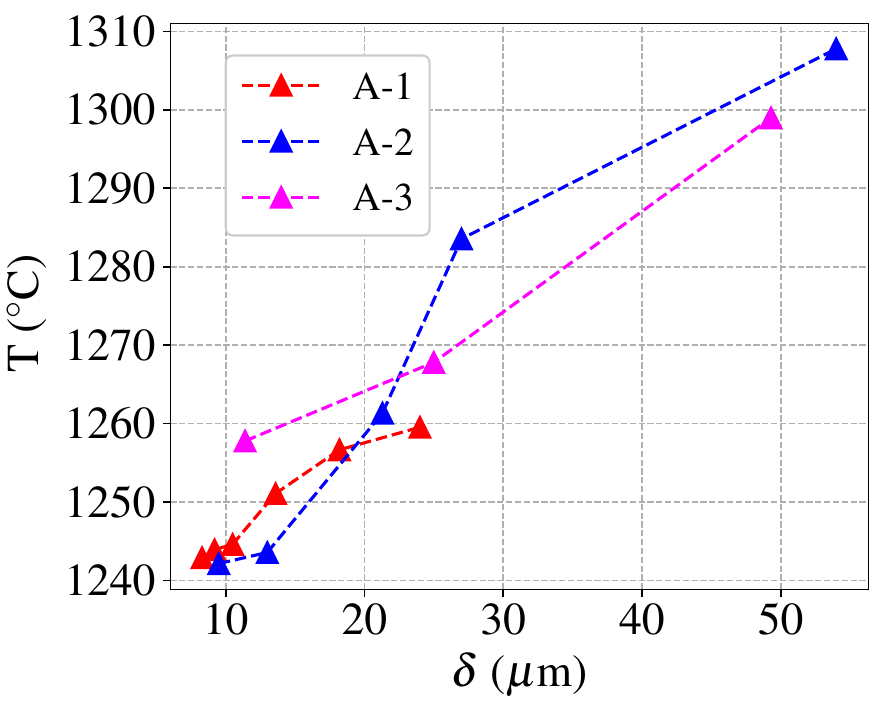}
	\caption{\rev{Maximum temperature sample A}}
  \label{fig:T_delta}
  \end{subfigure}
    \caption{
   Evolution of blisters under slow coooling rate: debonded diameter for  (a) sample A as a function of N and (b) sample B as a function of N,  (c) blister height for sample A as function of N  \rev{and (d)   maximum temperature at blisters from sample A as a function of blister height.} (full symbols denote perfect blisters, while empty symbols represent blisters with local or total spallation)
    \label{fig:blisterm_slow}}
\end{figure}

As outlined earlier, the LASDAM technique enables the monitoring of both in situ interfacial debonding and ex situ blistering \cite{Mahfouz:2023}. The evolution of decohesion diameter as a function of the number of cycles for BR-slow testing (A and B samples) is presented in Figures \ref{fig:DPC29} and \ref{fig:DPC25}. The progression of blisters in locations 2 and 3 exhibits remarkable similarity for both specimens, particularly concerning the number of cycles leading to the initiation of debonding, which occurs around 260 cycles for both specimens: The disparity between samples A and B is minimal, with a difference of no more than 10 cycles for location 3,  as illustrated in Figure \ref{fig:blisterm_slow}. Additionally, at N=500, no debonding is observed for blister \#1 in both samples. \rev{The number of debonding and spallation cycles for each blister is detailed in Table \ref{tab:specimendetails}}.

For sample A, profilometer measurements were conducted every 100 cycles, necessitating testing interruptions. Given the analogous debonding behavior observed in both samples, these interruptions are presumed to have a minimal impact on the overall damage rate. The evolution of blister height over the number of cycles for sample A is shown in Figure \ref{fig:delta_PC29}. The results indicate that, while the equivalent delaminated diameter initially remains relatively constant, the blister height undergoes a significant increase. This trend aligns with observations made during furnace thermal cycling, where blister evolution is initially characterized by the activation of a buckling mechanism until a point is reached where further growth of interfacial debonding is observed \cite{guipont:2019}. The height evolution presented in Figure \ref{fig:delta_PC29} also illustrates a threshold, approximately \SI{50}{\micro\meter}, beyond which spallation occurs. \rev{Additionally, Figure \ref{fig:T_delta} demonstrates a consistent increase in maximum temperature, plotted previously in Figure \ref{fig:T_cloques}, with increasing blister height, supporting the importance of analyzing local overheating in understanding blister evolution}.

As the crack tip temperatures associated with each blister location remain similar during the dwell at high temperature for both BR-slow and BR-fast conditions, the two loading conditions can be considered comparable in terms of interfacial oxidation. To eliminate the influence of initial blister dimensions, the relative debonding and relative blistering are used for comparison. These are described, respectively, as:
\begin{equation}
    \frac{D(N)-D(0)}{D(0)}
\end{equation}
and
\begin{equation}
    \frac{\delta(N)-\delta(0)}{\delta(0)},
\end{equation} 
where $D(N)$ and $D(0)$ represent the current and initial debonded diameters, and $\delta(N)$ and $\delta(0)$ represent the current and initial blister height respectively.

\begin{figure}[H]
    \centering
    \begin{subfigure}[b]{0.45\textwidth}
        \centering
        \includegraphics[width=\textwidth]{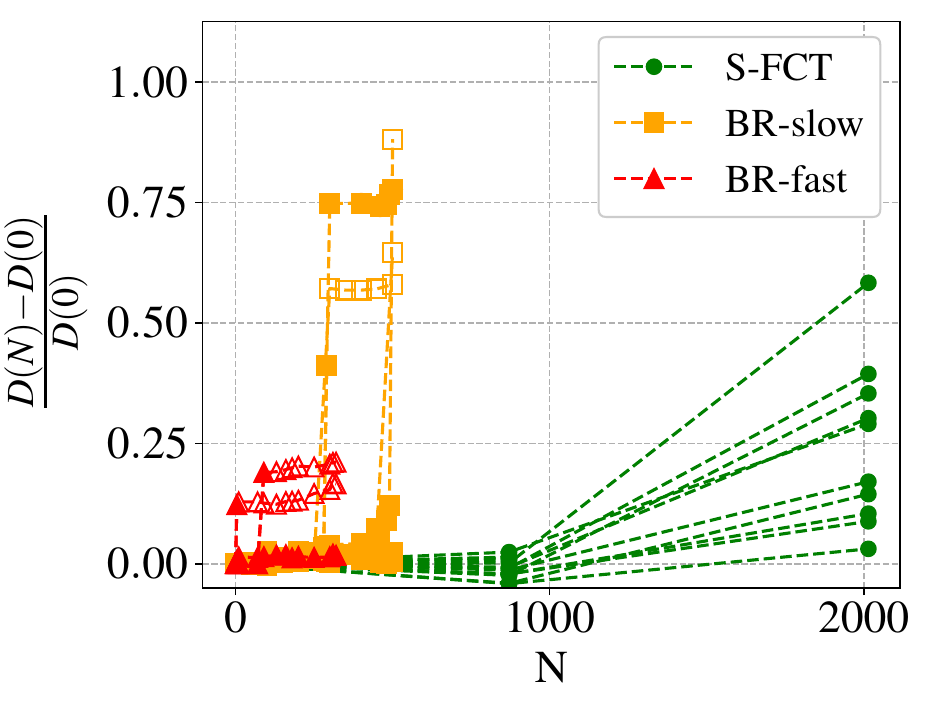}  
   \caption{relative diameter \label{fig:drel}}
   \end{subfigure}
    \begin{subfigure}[b]{0.45\textwidth}
        \centering
        \includegraphics[width=\textwidth]{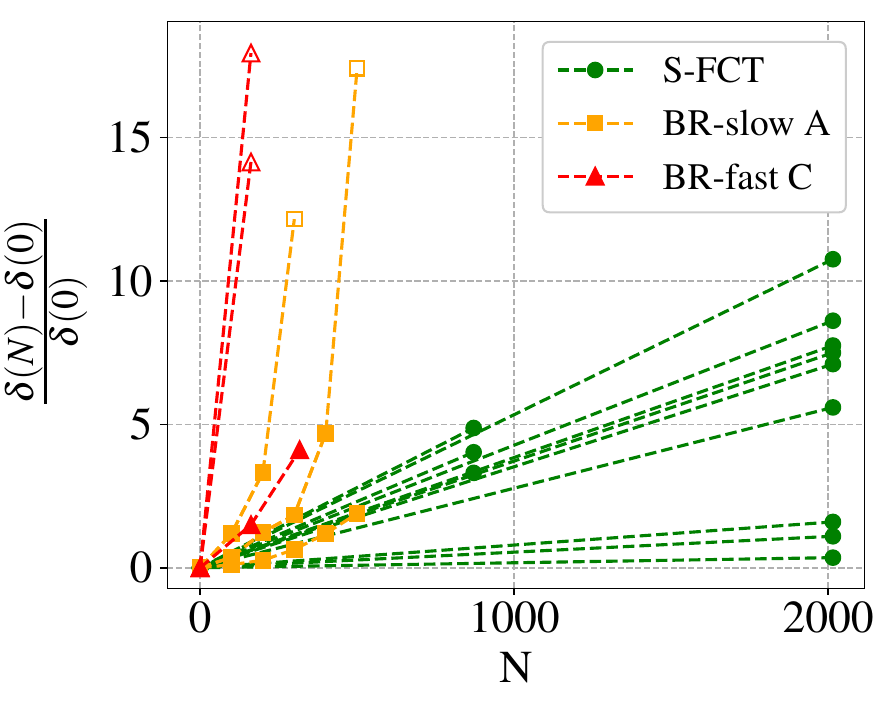}
   \caption{relative height \label{fig:hrel}}
    \end{subfigure}
    \caption{
    (a) Evolution of the relative debonded diameter with respect to the number of cycles; (b) Evolution of the relative blister height with respect to the number of cycles for both fast and slow cooling rates (full symbols denote perfect blisters, while empty symbols represent blisters with local or total spallation).}
  \label{fig:evolutionDam}
\end{figure}

Significantly accelerated macroscopic evolution kinetics are evident for both delamination and blistering in BR fast cycling compared to BR slow cycling, as shown in Figures \ref{fig:evolutionDam}(a) and (b), respectively. \rev{To clarify the role of thermal gradient on damage rate, S-FCT data were added to these plots, combining data from four different samples.  For three of these samples, three individual blisters are examined, while one other sample contains only one blister. To emphasize the blister condition, full symbols indicate blisters with no evidence of topcoat damage, while empty symbols indicate local cracking and/or spalling of the topcoat at the blister location}. In particular, local cracking or spallation manifests much earlier at the fast cooling rate, with the first local crack observed after only 10 thermal cycles and the second after 90 cycles, see Figure \ref{fig:evolutionDam} and Table \ref{tab:specimendetails} for values. In contrast, these events occur at later stages, specifically after 300 and 500 cycles, respectively, for the slow cooling rate.

\section{Discussion}
\subsection{Microstructure evolution for burner rig and FCT}

A previous study on the same system has provided data on microstructure evolution, particularly focusing on oxide thickness, denoted as h$_{ox}$, and phase transformation of the outer bond coat layer from $\beta$-(Ni,Pt)Al to $\gamma$'-Ni$_3$Al phase \cite{Maurel:2012b, soulignac2014prevision}. This was analyzed through the surface fraction (f$_s$) of phase $\gamma$'-Ni$_3$Al. The selected FCT cycle for this study, as described in \cite{Mahfouz:2023}, involves linear heating (and cooling) within 5 minutes, from 100 to \SI{1100}{\celsius}, with 5 minutes dwell at \SI{1100}{\celsius}.

\rev{Considering a similar cumulative time spent at maximum temperature, $t_{\textrm{HT}}$, the results of the BR experiment are very similar to those of the FCT regarding  oxide thickness, with the variation being relatively small compared to the inherent uncertainty in their quantification. However, the difference for phase transformation is more significant, suggesting a local lower temperature in the bond coat for burner rig compared to FCT. However, this point should be mitigated by the fact that this measurement is highly sensitive to the width of the analyzed area, e.g. as shown in previous results dealing with the same TBC system \cite{Maurel:2012f}. This suggests that the observed increase in blister evolution, both in height and debonding, is mainly controlled by a change in mechanical loading rather than a change in interfacial oxidation. The only significant difference in the microstructure evolution when comparing BR to FCT is the modification of the sintering at the center of the blister. This suggests, a priori, only a mechanical effect}.

\rev{A robust analysis of the temperature gradient is needed to more accurately assess the temperature reached at the crack tip during the BR cycle, by the way, the microstructure was observed only on the remaining adherent areas to avoid artifacts induced by local oxide spallation. This will be addressed in a future work}.

\subsection{Influence of temperature, temperature gradient and cooling rate on damage}
\label{sec:sample1}
 In a previous study on the TBC system investigated in this paper, LASDAM results were obtained under homogeneous temperature conditions throughout the system, using short Furnace Cycling Tests (FCT) with dwell times of 5 minutes \cite{Mahfouz:2023}. The same experimental methodology was then used: the interfacial defects were induced after an initial aging of 100 long FCTs (L-FCT) with a dwell time of 50 minutes at high temperature. Figures \ref{fig:drel} and \ref{fig:hrel} illustrate that \rev{both the damage rate associated with relative variations in debonded diameter and the stress buildup associated with relative variations in blister height are higher for burner rig (BR) cycling compared to S-FCT}.

\begin{figure}[H]
	\centering

 \includegraphics[width=0.55\textwidth]{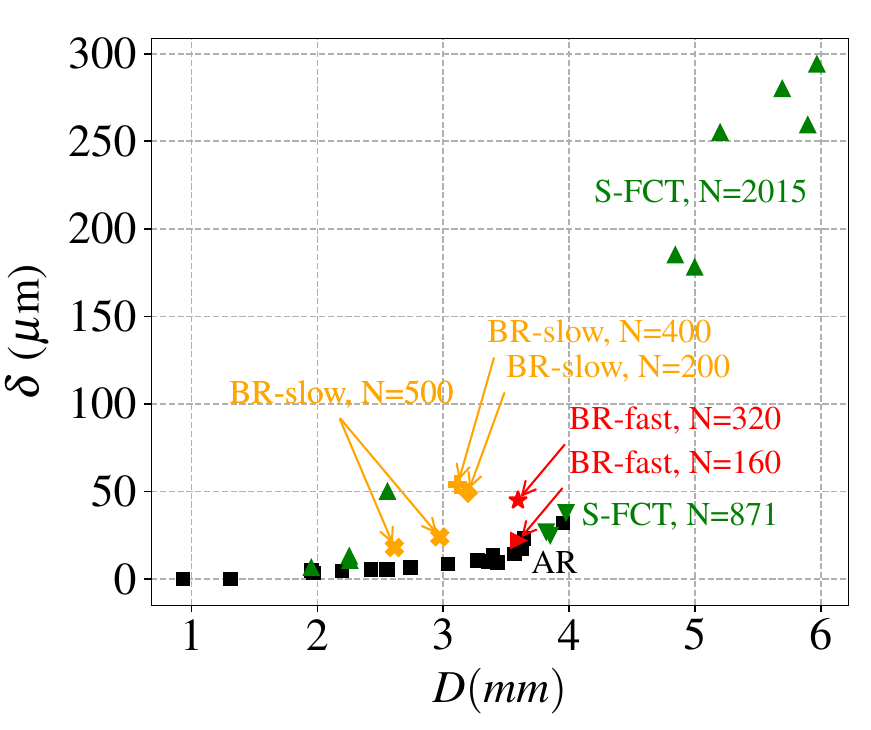}
	\caption{Relationship between blister height and debonded diameter for Short Furnace Cycling Test (S-FCT) fatigue (results from \cite{Mahfouz:2023}) and Burner Rig (BR) experiment (current study, with maximum height measured before spallation)}. \label{fig:comparison}
	
\end{figure}

Another aspect of the analysis is to evaluate the height-diameter relationship, which serves as an indicator of the stress state within the blister. In this context, for a given diameter, a higher blister height corresponds to increased stresses within the blister \cite{mcdonald:2010,Mahfouz:2023}. In the as-received (AR) condition, corresponding to the LASAT defect introduced after 100 preliminary L-FCTs, indicated by the black symbols in Figure \ref{fig:comparison}, the maximum height is approximately \SI{40}{\micro\meter} for a diameter of \SI{4}{mm}. No noticeable difference in height compared to the AR state is observed after 160 cycles of BR fast cycling. However, for a higher number of cycles, a trend becomes apparent indicating increased stress in the system for both BR-fast and BR-slow conditions.
The blister height obtained under BR conditions approaches that achieved after 2015 S-FCT cycles, as indicated by the green triangle symbols in Figure \ref{fig:comparison}. This further demonstrates that BR testing is more detrimental than S-FCT, as the stress build-up is higher in the former than in the latter, \rev{in consideration of the lower number of cycles}.

\rev{On the other hand, considering the similar oxidation and phase transformation rates in BR and FCT conditions, it could be assumed that the average bond coat temperature at the crack tip was similar for both test conditions. This point is of high interest because it has been shown in the LASDAM experiment that the BC plasticity plays a key role in determining the applied torque to the blister \cite{Mahfouz:2023}: the higher the plasticity in the BC, the higher the blistering and damage rate. Thus, the local behavior of the BC might be comparable}.

\subsection{Impact of temperature, temperature gradient and cooling rate on energy}
The previous analysis is entirely focused on experimental outcomes. It is worth exploring whether state of the art approaches of mechanical loading analysis would yield consistent conclusions. %It is worth asking whether the temperature conditions tested are consistent with the state of the art in mechanical loading. 
For sake of simplicity, we initially describe the system's energy without considering the blister but incorporating the experimentally measured temperature gradient across adherent coating. 

The first selected model is proposed by Levi et al. in Ref. \cite{levi2012environmental}. The authors evaluate the elastic energy, $U$, stored in the top layer, assuming plane-strain conditions. The interest of the model is  its sensitivity to the temperature gradient across the interface and the temperature drop on cooling, for both the top coat and  the substrate. The model follows:
\begin{equation}\label{eq:energy}
    U(t)=\frac{E(t) h (1+\nu)}{2(1-\nu)}\Big\{(\Delta\alpha\Delta T_{sub})^2-(\Delta\alpha\Delta T_{sub})(\alpha\Delta T_{sur/sub})+\frac{1}{3}(\alpha\Delta T_{sur/sub})^2\Big\}
\end{equation}
where E, $\nu$, and $\alpha$ are the Young's modulus, Poisson's coefficient, and CTE of the TC layer, respectively. The thickness of the TC is denoted $h$. The CTE mismatch between the substrate and the TC is denoted $\Delta\alpha$. The temperature gradient $\Delta T_{sur/sub}$ is defined as the difference between the temperature of the surface and that of the substrate, and the cooling $\Delta T_{sub}$ as the substrate temperature drop from the maximum temperature during dwell time.  The model assumes an equi-biaxial stress state function of the analyzed point in the ceramic thickness $y$ :
\begin{equation}\label{eq:sigy}
\sigma(y)=\frac{E }{1-\nu}\big(-\Delta\alpha\Delta T_{sub}+(y/h)\alpha\Delta T_{sur/sub}\big)
\end{equation}

\begin{figure}[H]
    \centering
    \begin{subfigure}[b]{0.45\textwidth}
    \centering
        \includegraphics[width=\textwidth]{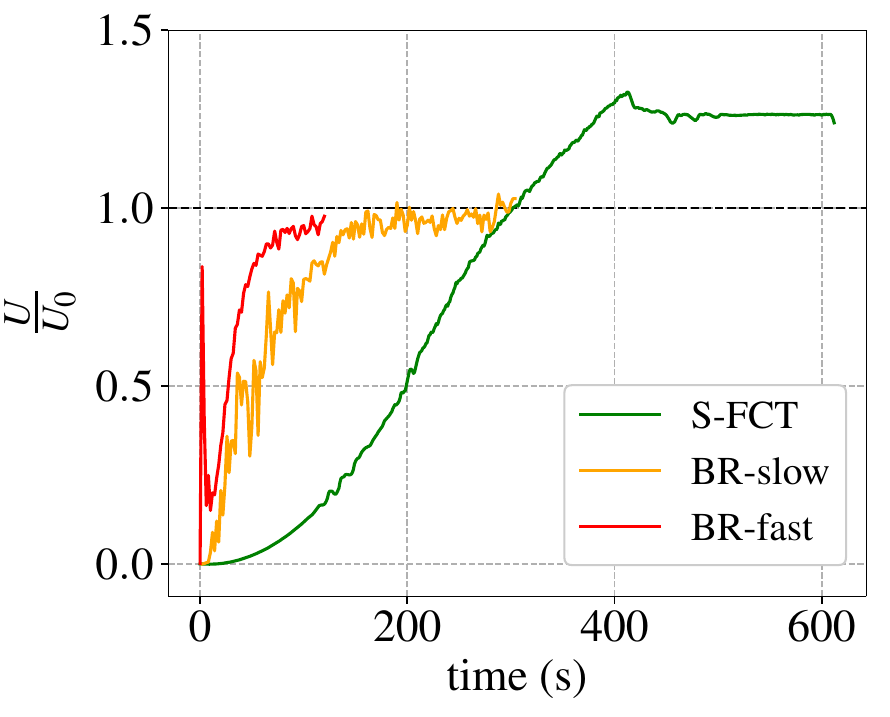} 
   \caption{Using model from \cite{levi2012environmental}}.
   \label{fig:ERR_Levi}
   \end{subfigure}
    \begin{subfigure}[b]{0.45\textwidth}
    \centering
        \includegraphics[width=\textwidth]{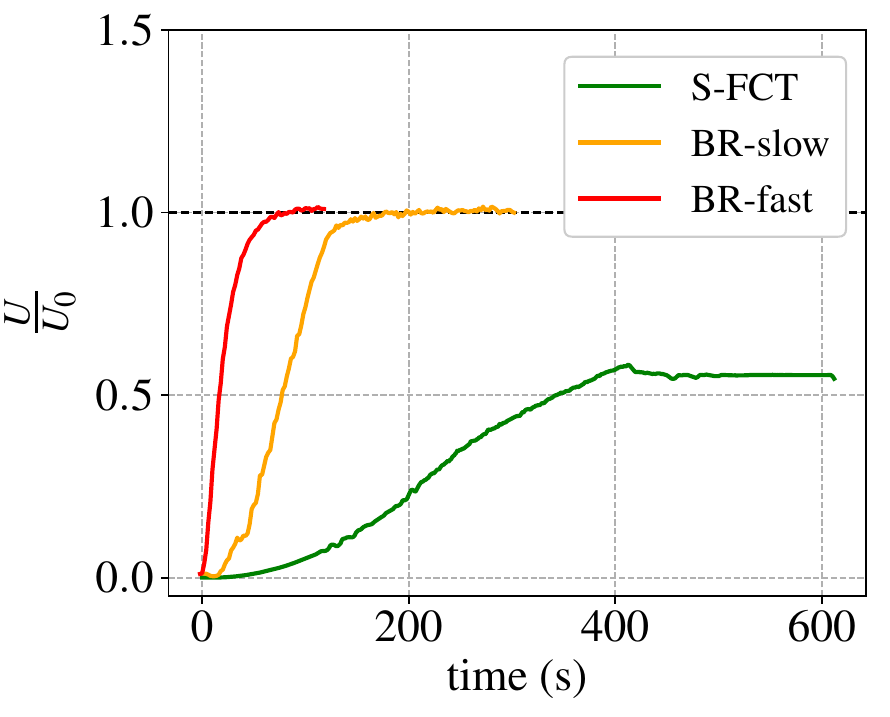}   
   \caption{Using model from \cite{vassen2009}}. 
   \label{fig:ERR_Vassen}
   
   \end{subfigure}
   \caption{Normalized elastic energy per unit area in the coating using experimental measurements of thermal condition on cooling}
  \label{fig:ERR}
\end{figure}

Using this model, the temperature evolution during cooling from experimental data is used to evaluate BR-slow and BR-fast conditions, as shown in Figure \ref{fig:cooling_blisters}. To provide a benchmark, S-FCT cycle is examined, assuming a linear cooling rate of \SI{3.3}{K/s} and no gradient (i.e., $\Delta T_{sur/sub}=0$ in equation \ref{eq:energy}). The energy is normalized based on its value obtained for BR at the minimum temperature shown in Figure \ref{fig:ERR_Levi} to facilitate comparative analysis.

The normalized energy release rate curves suggest a faster reaching of the maximum energy level of the system at higher cooling rates. The asymptotic values of the energy release rate at the minimum temperature are identical for both BR-slow and BR-fast conditions, but a significant transient difference is observed: BR-slow shows a monotonic energy increase, while BR-fast shows a peak in the first few seconds of cooling. As demonstrated in \cite{levi2012environmental}, the cooling of the coating relative to the substrate causes tensile stresses in the upper layer, resulting in peak energy release rates which may exceed those observed in the cooled state. This tensile stress phenomenon was further validated by finite element modeling of thermal gradient cycling in \cite{Mahfouz:2023}. Surprisingly, S-FCT achieves a higher steady-state energy than the burner rig conditions, contradicting the experimental results. \rev{The model proposed by Levi et al. \cite{levi2012environmental} explains that the energy release rate in coatings subjected to thermal cycling is influenced by two main factors: the differential contraction arising from the mismatch in thermal expansion coefficients between the substrate and the coating, and the self-constriction of the coating due to cooling at its surface. In scenarios where the cooling rate is slow for thermal gradient cycling conditions, the gradient’s effect can help alleviate the strains and stresses experienced by the coating, potentially lowering the energy release rate}. This inconsistency with experimental results suggests a potential sensitivity of damage to cooling rates, implying that faster energy input may result in greater damage.

The curves in Figure \ref{fig:ERR_Levi} were contrasted with those generated using Va{\ss}en et al.'s model from \cite{vassen2009}. The latter estimates the energy release rate stored in the coating by integrating the in-plane stresses across the thickness of the spalling coating and incorporating these values into the energy release rate of a straight interface crack. 
\rev{This model follows:
\begin{equation}
    U(t)=\tau\frac{1-\nu^2}{2(1-\nu)^2}\int_y^{h}E(t,x)\Big\{\Delta T_{sur/sub}\alpha+\Delta T_{sub}\Delta\alpha\Big\}^2 dx
\end{equation}}
\rev{where $\tau$ is a parameter fitted to match the observed time to spallation, other parameters being consistent with the notations used for Levi's model in equation \ref{eq:energy}}.

The resulting normalized energy release rate curves are shown in Figure \ref{fig:ERR_Vassen}.

Comparing Levi's model with Va{\ss}en et al.'s \cite{vassen2009}, the latter presents a lower stabilized energy release rate for furnace cycling, aligning better with experimental results. However, it fails to identify an energy peak for BR-fast at the start of cooling.

\begin{figure}[H]
   \begin{subfigure}[b]{0.45\textwidth}
   \centering
        \includegraphics[width=\textwidth]{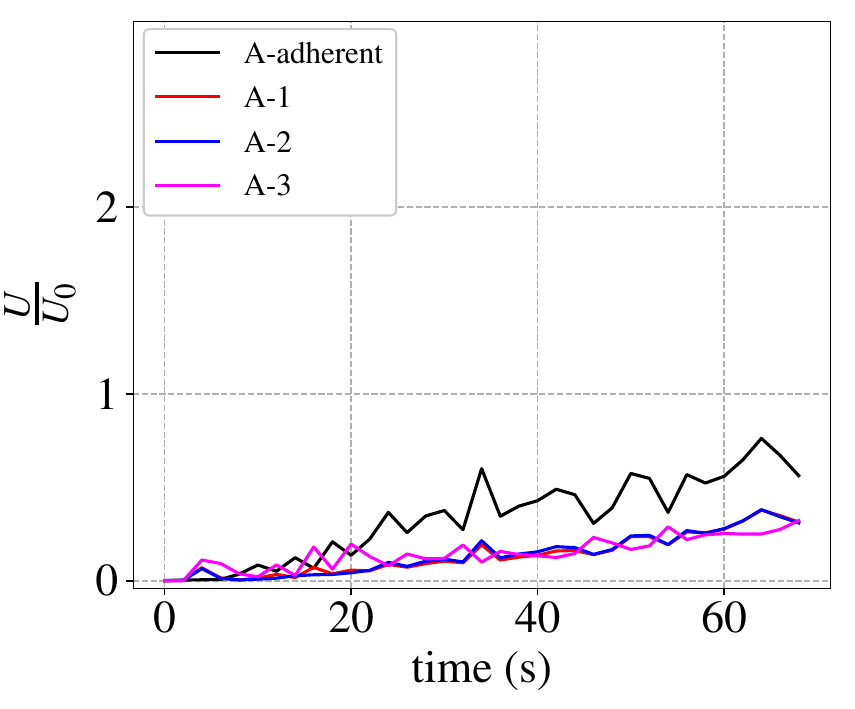}   
   \caption{BR-slow (A) at N = 250}
   \label{fig:N=250}
   \end{subfigure}
    \begin{subfigure}[b]{0.45\textwidth}
    \centering
        \includegraphics[width=\textwidth]{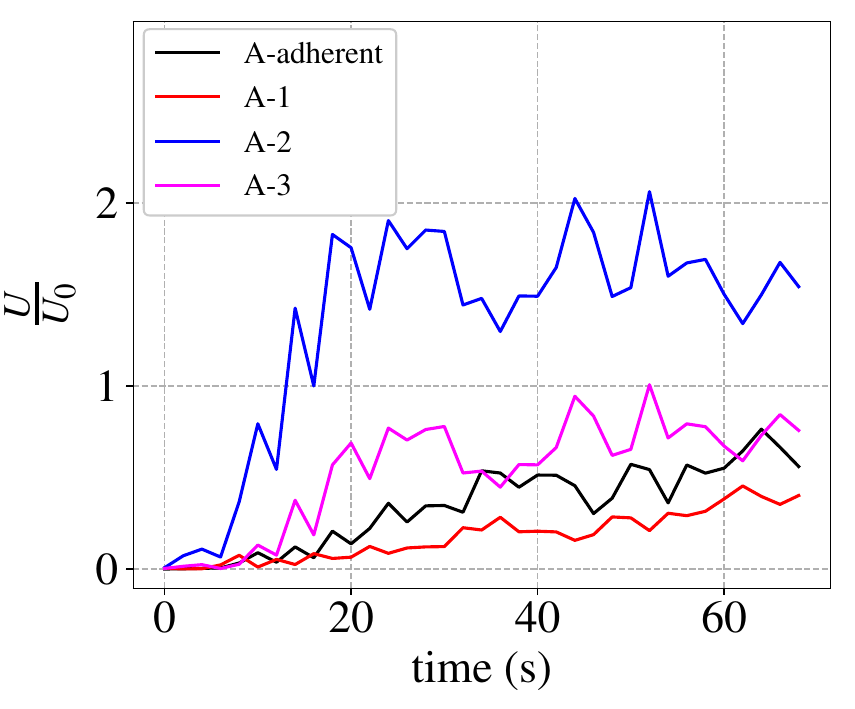}
   \caption{BR-slow (A) at N = 500}
   \label{fig:N=500}
    \end{subfigure}
    \caption{Normalized elastic energy per unit area in the coating, based on Levi's model eq. \ref{eq:energy} \cite{levi2012environmental}}
  \label{fig:U_blistersPC29}
\end{figure}

The thermal loading analyzed in section \ref{sec:dwell_maxT} shows that much higher temperatures were reached at the blister sites, with these local temperatures increasing during cycling as a result of blistering. Considering the temperature profiles extracted at the maximum temperatures at the blister locations from infrared thermography measurements during cooling to \SI{400}{\celsius}, normalized energy curves \rev{based on Levi's model, eq. \ref{eq:energy}}, are plotted in Figure \ref{fig:U_blistersPC29} for Sample A tested with the BR slow cycle, at cycle 250 and cycle 500. This approach is a rough approximation of the mechanical state as Levi et al.'s model \cite{levi2012environmental}, subsequently used, ignores the change in stress state in the presence of a blister. The aim is to evaluate the trend in terms of the effect of overheating on the mechanical state. This analysis is subsequently only qualitative.

\begin{figure}[H]
	\centering
 \includegraphics[width=0.55\textwidth]{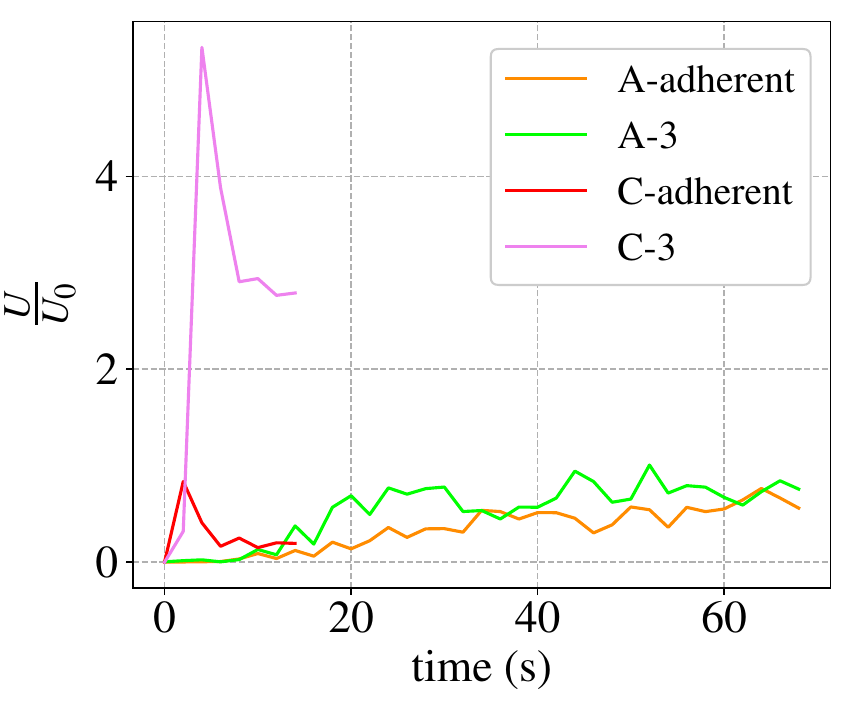}
	\caption{At blisters for BR-fast and BR-slow during cooling down to \SI{400}{\celsius}, based on Levi's model eq. \ref{eq:energy} \cite{levi2012environmental}}. \label{fig:U_T}
\end{figure}

Observing the cooling process at cycle 250, it is noteworthy that the elastic energy release rate calculated at the adherent top coat area exceeds that at the blister sites, Figure \ref{fig:N=250}. As explained earlier, according to Levi et al.’s model, a mild thermal gradient throughout the top coat can be beneficial and potentially lower the energy release rate. This is evident in the case of blisters A-1, A-2 and A-3 compared to the adherent top coat at N=250. However, as the gradient increases for A-2 and A-3, it amplifies the energy release rate compared to the adherent top coat. In fact, at cycle 500, the elastic energy release rate has escalated for blisters A-2 and A-3, with A-2 showing higher energy values, Figure \ref{fig:N=500}. \revn{ Figure \ref{fig:U_T} takes into account comparable maximum temperatures at the blister sites for BR-slow and BR-fast, and cooling down to 400°C, which occurs within approximately 20 seconds under fast cooling rates, and around 70 seconds under slow cooling rates. The normalized energy curves show exceptionally high energy values for blister C-3, which peaks at a normalized value of about 5}.

Thus, limiting the analysis to the adherent temperature alone fails to provide strong insights. However, considering the actual thermal loading at blister locations reveals higher steady-state energy during both peak and minimum temperatures in burner rig cycling compared to furnace cycling.

In conclusion, both models provide valuable insights into both asymptotic and transient aspects. The analysis further emphasizes that the cooling rate transient significantly influences the mechanical state in the TBC. 

 Local overheating of the blister dictates the steady-state behavior, highlighting the strong influence of the blister on the mechanical loading of the TBC. These results motivate numerical solutions to evaluate a more realistic mechanical state, taking into account both the temperature gradient and the mechanical evolution of the system, as well as the modification of the blister dimensions during thermal cycling. \rev{The strong correlation between local maximum temperature and blister height is a key factor in the local TBC stress as shown in Figure \ref{fig:T_delta} and should be considered in further studies. The coupling between oxidation rate, growth strain and BC creep has been established as a crucial parameter controlling the mechanical state of the TBC considering the homogeneous temperature in the FCT \cite{Mahfouz:2023}. This drastic temperature gradient should therefore play a key role in the local loading, i.e. the stress intensity factor or other driving force controlling the damage rate}. 
 
\rev{Last but not least, the measurement of temperature in FCT specimens is not as precise compared to the in-depth analysis of BR experiments proposed in this study: in FCT specimens, thermocouples do not take into account either the temperature gradient in the systems or the thermal inertia of the specimen, which could affect the above analysis. Furthermore, a more accurate temperature measurement should be considered for each experimental condition in order to conclude on the transient aspects in the temperature evolution and the subsequent impact on the TBC damage, this possibility being also offered by the LASDAM method.}

\section{Conclusions}
This study demonstrates the efficiency of the LASDAM method for testing complex configurations, accounting for thermal gradients, and notably, avoiding edge effects. In essence, damage localization is initiated by the presence of the initial blister, enabling robust monitoring of the damage rate at that specific location. The harsh environment induced by the burner rig condition has been thoroughly characterized through surface analysis of temperature. The in-depth characterization of the damage rate highlights the substantial impact of temperature and, more crucially, cooling rate on damage evolution. The through-thickness temperature gradient is more challenging to analyze, but a comparison of thermocouple temperature measurements within the volume of the substrate and surface temperature measurements indicates that the inversion of the temperature gradient during cooling for a higher cooling rate must drive the higher damage rate compared to a lower cooling rate.

For the first time, it is revealed that the top coat layer experiences very high local temperatures due to the blister and the associated air gap between TC and oxide. This scenario mirrors real component applications where delamination precedes final spallation. This local temperature increase is observed both through in situ temperature field measurement with IRT and the sintering of the TC: proximity to the center of the blister corresponds to higher temperatures and increased sintering.

Comparison with furnace thermal cycling is difficult due to uncertainties in local interface temperature. However, the cooling rate in burner rig testing is typically higher than in FCT testing. As demonstrated for APS coatings, both the higher local temperature and the higher cooling rate have a synergistic effect on the increased damage rate observed under burner rig conditions compared to FCT conditions. 
\rev{In summary, the following findings were thoroughly addressed in this study:}
\begin{itemize}
	\item The introduction of pre-defect using the LASDAM method allows in-situ monitoring of temperature and damage evolution;
	\item A fast cooling rate has been shown to increase the damage rate compared to a slow cooling rate;
	\item Local overheating is a direct function of the air gap between the blister and the metallic bond layer. \revn{SEM analysis shows that the overheating is high enough to promote sintering of the TC layer at the top of the blister, while the adherent TC remains unaffected};
	\item Sintering, oxide thickness and phase transformation tend to justify that the temperature at the crack tip associated with the artificial blister is consistent with the adherent areas of the TC, thus limiting the artifact associated with overheating;
	\item However, the mechanical state of the blister is significantly altered by this overheating;
	\item Models that consider the elastic energy stored in the TC layer underscore the critical role of fast cooling rates, particularly evident in the Levi’s model, where thermal transients result in a sudden increase of elastic energy at high temperature;
	\item Comparisons with standard furnace cycling tests are limited due to data constraints.
\end{itemize}
\rev{The significant influence of the thermal transient has been clearly demonstrated using the LASDAM technique and should be further strengthened in the FCT analysis}. These aspects should be addressed in future publications through a combination of finite element analysis and correlation with experimental temperature measurements. 
In addition, the LASDAM method has the potential for multiple shock analysis, allowing further exploration of in-plane gradients, which should be developed in future work.

\section*{References}

%% If you have bibdatabase file and want bibtex to generate the
%% bibitems, please use
%%
 %\bibliographystyle{elsarticle-num} 
 %\bibliography{refs,gradbib}

 \end{document}